\documentclass[journal]{IEEEtran}
\usepackage{epsfig} 
\usepackage{amsmath} 
\usepackage{amssymb}  
\usepackage{epstopdf}
\usepackage{subfigure}  
\usepackage{cite}
\usepackage{balance}
\usepackage{hyperref}
\usepackage{color}
\usepackage{multirow}
\usepackage{booktabs}
\usepackage{algorithm}
\usepackage{booktabs}

\newtheorem{ass}{Assumption}

\newtheorem{rem}{Remark}

\newtheorem{lemma}{Lemma}
\allowdisplaybreaks

\graphicspath{{./figures/}}

\ifCLASSINFOpdf
\else
\fi

\begin{document}
\title{
A Unified Model for Active Battery Equalization Systems
} 

\author{Quan~Ouyang,~\IEEEmembership{Member,~IEEE,} Nourallah Ghaeminezhad, Yang Li,~\IEEEmembership{Senior Member,~IEEE,}  
Torsten Wik,~\IEEEmembership{Member,~IEEE,}  Changfu~Zou,~\IEEEmembership{Senior Member,~IEEE}
\thanks{This work was supported by  Marie Skłodowska-Curie Actions Postdoctoral Fellowships under the Horizon Europe programme (Grant No. 101067291). (Corresponding author: Changfu Zou)}


\thanks{Quan~Ouyang, Yang~Li, Torsten~Wik, and Changfu~Zou are with the Department of Electrical Engineering, Chalmers University of Technology, 41296 Gothenburg, Sweden. (e-mail: quano@chalmers.se; yangli@ieee.org; tw@chalmers.se; changfu.zou@chalmers.se)}

\thanks{Nourallah~Ghaeminezhad is with the College of Automation Engineering, Nanjing University of Aeronautics and Astronautics, Nanjing 211100, China. (e-mail: ghaemi@nuaa.edu.cn)}
}

\markboth{}
{Shell \MakeLowercase{\textit{et al.}}: Bare Demo of IEEEtran.cls for IEEE Journals}


\maketitle

\begin{abstract}	
Lithium-ion battery packs demand effective active equalization systems to enhance their usable capacity and lifetime. Despite numerous topologies and control schemes proposed in the literature, conducting quantitative analyses, comprehensive comparisons, and systematic optimization of their performance remains challenging due to the absence of a unified mathematical model at the pack level. To address this gap, we introduce a novel, hypergraph-based approach to establish the first unified model for various active battery equalization systems. This model reveals the intrinsic relationship between battery cells and equalizers by representing them as the vertices and hyperedges of hypergraphs, respectively. With the developed model, we identify the necessary condition for all equalization systems to achieve balance through controllability analysis, offering valuable insights for selecting the number of equalizers. Moreover, we prove that the battery equalization time is inversely correlated with the second smallest eigenvalue of the hypergraph's Laplacian matrix of each equalization system. This significantly simplifies the selection and optimized design of equalization systems, obviating the need for extensive experiments or simulations to derive the equalization time. Illustrative results demonstrate the efficiency of the proposed model and validate our findings.
\end{abstract}

\begin{IEEEkeywords}
Active battery equalization, unified model, hypergraph, equalization time estimation.
\end{IEEEkeywords}

\IEEEpeerreviewmaketitle

\section*{Nomenclature} \addcontentsline{toc}{section}{Nomenclature}
\begin{IEEEdescription}[\IEEEusemathlabelsep\IEEEsetlabelwidth{$V_1,V_2,V_3$}]

\item[$A$] State-transition matrix of the closed system.
\item[$B_i$] Vertice corresponding to the battery cell $i$.
\item[$b$] Number of cells in each battery module.
\item[$C$] Incidence matrix.
\item[$c_l$] Incidence vector corresponding to $e_l$.
\item[$c_{p,l}$] Element of the incidence matrix $C$.
\item[$D$, $d$] System matrices.
\item[$d_s$] Smallest diagonal elment of $D$.
\item[$e_l$] Hyperedge corresponding to the equalzier $l$.
\item[$G$] Hypergraph corresponding to the active equalization system.
\item[$H(e_l)$] Head of the hyperedge $e_l$.
\item[$I_{ec_{i,l}}$]   Equalization current of cell $i$  through the CC equalizer $l$.
\item[$I_{em_{i,l}}$] Equalization current of cell $i$  through the MM equalizer $l$.
\item[$I_{ep_{i,l}}$] Equalization current of cell $i$ through the CPC equalizer $l$.
\item[$I_{ecm_{i,l}}$] Equalization current of cell $i$ through
the CMC equalizer $l$.
\item[$I_{e_l}$] Equalization current vector for all cells through equalizer $l$.
\item[$I_{eq}$] Total equalization current vector of all cells.
\item[$I_{eq_i}$]  Total equalizing current of cell $i$.

\item[$I_{{ec}_l}$] Directed equalization current provided by the CC equalizer $l$.
\item[$I_{em_l}$] Directed equalization current provided by the MM equalizer $l$.
\item[$I_{ep_{l}}$] Directed equalization current on the cell side provided by the  CPC equalizer $l$.
\item[$I_{ecm_l}$] Directed equalization current on the cell side provided by the CMC equalizer $l$.
\item[$I_{l}$] Directed equalization current provided by equalizer $l$ in any type of active equalization system.
\item[$\bar{I}_{ec_l}$] Magnitude of the equalization current provided by the CC equalizer $l$.
\item[$\bar{I}_{em_l}$] Magnitude of equalization current provided by the MM equalizer $l$.
\item[$\bar{I}_{ep_l}$ ] Magnitude of the equalization current on the cell side provided by the CPC equalizer $l$.
\item[$\bar{I}_{ecm_l}$] Magnitude of the equalization current on the cell side provided by the CMC equalizer $l$.

\item[$I_s$]  Current of the battery pack through the external power source or load.

\item[$K$] Control gain matrix.
\item[$k_s$] Smallest diagonal elment of $K$.
\item[$k$] Discrete time index.
\item[$L$] System matrix for controllability analysis.

\item[$m$] Number of modules in the battery pack.
\item[$n$] Number of cells in the battery pack.
\item[$n_e$] Total number of equalizers.

\item[$Q_i$]   Ampere-hour capacity of cell $i$.
\item[$SOC$] State-of-charge of the battery cell.
\item[$\bar{SOC}_{m_i}$] Average SOC of the battery module $i$.
\item[$\bar{SOC}_{P}$] Average SOC of the battery pack. 
\item[$s$]  System state vector for controllability analysis.

\item[$T(e_l)$] Tail of the hyperedge $e_l$.
\item[$T_e$] Equalization time. 

\item[$T_0$] Sampling period.
\item[$V_{B}$] Terminal voltage of the battery cell.
\item[$w_{l_h}$, $w_{l_t}$] Weights corresponding to tail and heads of the hyperedge.
\item[$x$, $u$] State vector and control variable vector of the battery equalization system.

\item[$\alpha_l$] Energy transfer efficiency of the $l$-th CC equalizer.

\item[$\eta$] Coulombic efficiency.

\item[$\epsilon$] Upper bound of the SOC imbalance tolerance.

\end{IEEEdescription}

\section{Introduction}
The demand for high-performance lithium-ion battery systems has grown exponentially in recent years with their widespread adoption in electric vehicles, energy storage systems, and portable electronic devices. Since a single lithium-ion battery cell's voltage is limited to $2.4$--$4.2$\rm{V} due to its inherent electrochemical characteristics, a large number of cells are usually connected in series and parallel in a battery pack to meet the high-voltage and large-capacity requirement for practical applications \cite{FENG2019109464}. However, cell imbalances caused by inhomogeneous conditions and manufacture variations result in insufficient energy use,  accelerated capacity degradation, and even safety hazards of the entire battery pack \cite{CHEN2022100025, Chen23, SHEN2023100109}.
Thus, the battery pack necessitates an active battery equalization system that can transfer energy from cells with high state-of-charge (SOC) to cells with low SOC. By this means, the cells can be charged and discharged more uniformly, thereby maximizing the available pack capacity and prolonging its useful lifetime \cite{HAN2017166, 9751217}.

Various active battery equalization systems have been developed to ensure all cells' SOCs converge to the same level \cite{Javier2014, 7762931}. In particular, hardware circuits of the active battery equalizers have been extensively studied, which include but are not limited to the double-tiered switching capacitor \cite{4451076}, multiwinding transformer \cite{7277091}, isolated modified buck-boost converter \cite{7995102}, bi-directional flyback converter \cite{9807338}. Additionally, there are also numerous effective balancing control methods, such as those based on neuro-fuzzy control \cite{Nguyen14}, quasi-sliding mode control \cite{8030126}, model predictive control (MPC) \cite{8337096}, and two-layer MPC \cite{9477598}.  


Compared to these achievements in hardware circuits and control methods, much less attention has been focused on quantitative analysis of different equalization systems at the pack level, which, however, is crucial in performance evaluation and comparison of different equalization structures. As one of the few examples, Chen {\it et al.} quantitatively compared the system-level performance of three types of active equalization structures, i.e., the series-based, layer-based, and module-based cell-cell (CC), using three individually developed models \cite{6832590}. 
As an extension, the mathematical models of seven balancing structures were comprehensively reviewed in \cite{9328206}, with additional consideration of cell-pack-cell (CPC), module-based CPC, directed CC, and mixed equalization systems. A similar work was carried out in \cite{9681274}, where nine analytical algorithms were proposed to estimate the equalization time of different active balancing structures. However, all the referred models 
are based on analyzing the total equalization currents received by each battery cell under specific structures of the equalization system, which will inevitably fail to capture the system characteristics at the battery pack level, rendering them incapable of mutual expansion and generalization. 

To bridge this identified research gap, this work, for the first time, develops a unified model for various active battery equalization systems using hypergraphs. By representing the equalizers as hyperedges and the cells as vertices, this method can reveal the intrinsic relationship between cells and equalizers.
Based on the developed unified model, the minimum required number of equalizers to complete the balancing task is determined through controllability analysis, and the equalization time is proved to be related to the Laplacian matrix of the hypergraph of the equalization system.

The major contributions of this paper are summarized in the following.

\begin{enumerate}
\item Different from the existing methods of separately modeling each equalization system, a unified mathematical model based on hypergraphs for various existing equalization systems is developed. This model provides a holistic, convenient, and concise format that applies to many different battery equalization systems, facilitating a range of model-based applications, e.g., analysis, comparison, optimization, and control design.

\item Based on controllability analysis, the necessary condition for equalization systems to achieve battery balancing is determined, which can provide guidance for finding the minimum number of equalizers required for a given equalization system.

\item Based on the unified model developed, the battery equalization time is shown to negatively correlate with the second smallest eigenvalue of the Laplacian matrix of the equalization system, which is demonstrated by the Monte Carlo method.

\end{enumerate}

The remainder of this paper is organized as follows. An overview of state-of-the-art active battery equalization systems is provided in Section~\uppercase\expandafter{\romannumeral2}. The unified model using hypergraphs is proposed in Section~\uppercase\expandafter{\romannumeral3} for all the considered active equalization systems. Based on the developed model, Section~\uppercase\expandafter{\romannumeral4} rigorously analyzes and comprehensively compares the performance of different equalization systems. Detailed illustrative examples and discussions are provided in Section~\uppercase\expandafter{\romannumeral5}, followed by conclusions in Section~\uppercase\expandafter{\romannumeral6}.

\section{Active Equalization Systems and Their Analysis}\label{Sec:Overview}
In active battery equalization systems, a series of equalizers are commonly utilized to transfer extra energy from the cells with high SOC to those with low SOC. Given that the hardware circuits of active equalizers have been extensively studied \cite{Javier2014, 7762931, 4451076, 7277091, 7995102, 9807338}, this work primarily focuses on modeling and analyzing the equalization system structures at the pack level.

	\begin{figure*}[!htb]
	\begin{center}
		\includegraphics[width=0.95\linewidth, trim=10 15 8 3, clip]{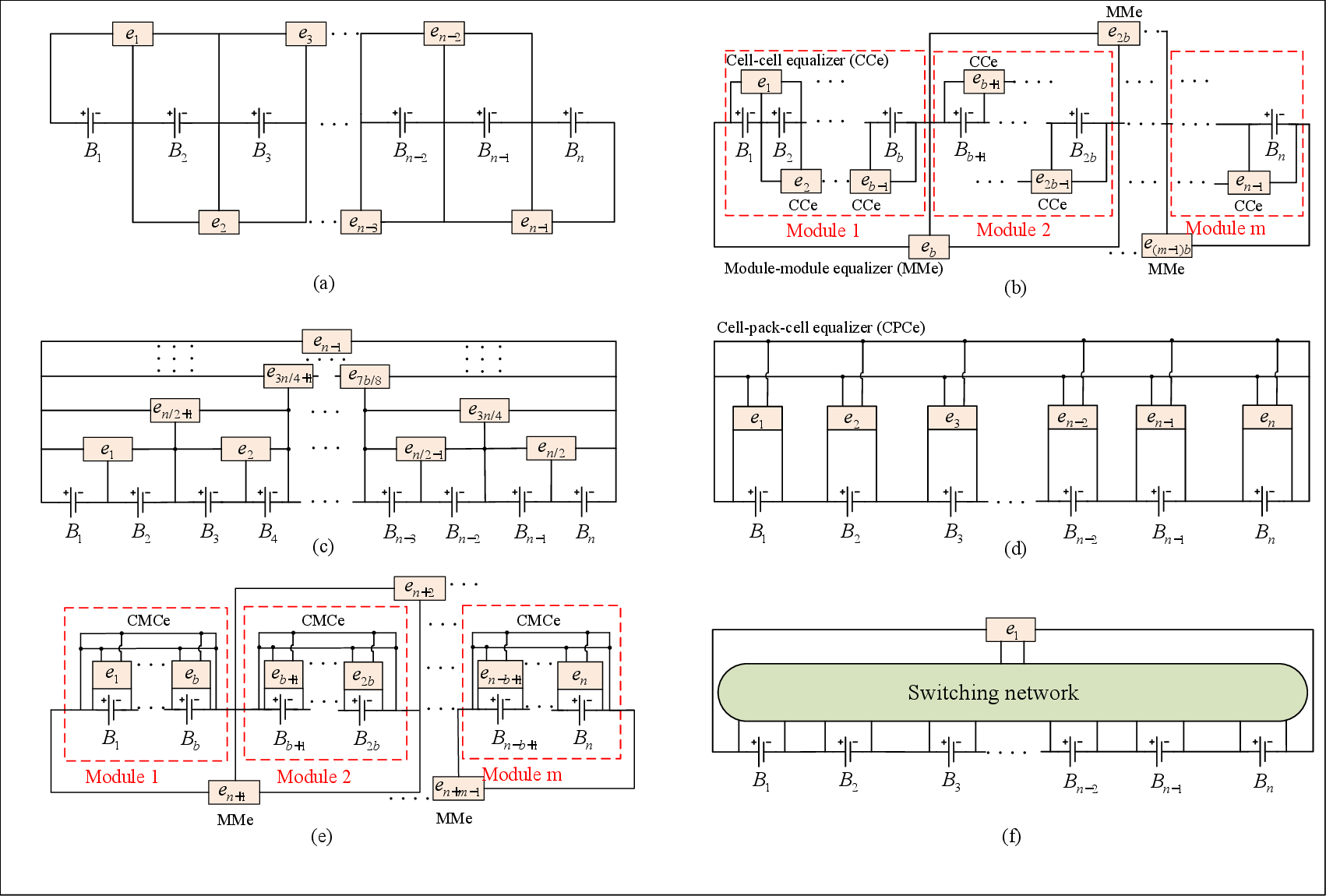}    
			\caption{The structures of six typical active equalization systems. Here, $B_i$ denotes battery $i$, and $e_i$ represents equalizer $i$. (a) Series-based CC, (b) module-based CC, (c) layer-based CC, (d) CPC, (e) module-based CPC, and (f) switch-based CPC equalization systems.}
		\label{model}
	\end{center}
\end{figure*}

\subsection{Overview of Battery Equalization Systems}\label{Sec:OverviewA}
Six typical active equalization systems are briefly introduced here, while more comprehensive reviews of battery equalization systems can be found in \cite{9328206} and \cite{9681274}.

\begin{enumerate}
\item Series-based CC equalization system. As illustrated in Fig.~\ref{model}(a), each CC equalizer connects a pair of adjacent cells and transfers energy between them. When the SOCs of all cell pairs are balanced, the cells' SOCs in the entire battery pack have reached the same level. For a battery pack with $n$ series-connected cells, $n-1$ CC equalizers are required.
	
\item  Module-based CC equalization system. As illustrated in Fig.~\ref{model}(b), the battery pack is divided into $m$ series-connected modules, where each battery module contains  $b=\frac{n}{m}$ series-connected cells. It enables balancing operations similar to the series-based CC equalization system within each module, while also taking into account neighboring modules for balancing. This equalization system includes $m-1$ module-module (MM) equalizers and $n-m$ CC equalizers.
	
\item Layer-based CC equalization system. As illustrated in Fig.~\ref{model}(c), this system employs a binary-tree-based structure, with multiple layers containing different numbers of equalizers according to the layer number. In each layer, two adjacent cells/modules are connected to the CC/MM equalizers to achieve balancing, following the same strategy as used in the module-based CC equalization system.
	
\item CPC equalization system. As illustrated in Fig.~\ref{model}(d), there exist $n$ CPC equalizers for the battery pack with $n$ series-connected cells, where each CPC equalizer connects each cell and the entire battery pack. Energy is transferred between the cell and the battery pack by the CPC equalizer when the cell's SOC is different from the average SOC of the battery pack.

\item Module-based CPC equalization system. As illustrated in Fig.~\ref{model}(e), the battery pack is divided into $m$ modules, similar to the module-based CC equalization system. In each module, cell-module-cell (CMC) equalizers are utilized to perform the same operation as in the CPC equalization system to achieve equalization. Thus, $n$ CMC equalizers are needed, and $m-1$ MM equalizers are utilized for balancing all the battery modules.
	
\item Switch-based CPC equalization system. As illustrated in Fig.~\ref{model}(f), all battery cells share one CPC equalizer, and a switching network controls which cell is connected to the equalizer. This structure allows bidirectional energy transfer between any selected cell and the battery pack. 	

\end{enumerate}
To the best of our knowledge, in the existing work comparing different equalization systems, such as \cite{8445645, 9328206, 9681274}, the different battery equalization systems are modeled separately, and then individually tailored models are utilized to simulate and evaluate the performance of each system. This is mainly because most models of active battery equalization were developed by analyzing the total equalization currents received by each battery cell under a specific equalization structure. In general, these models cannot capture the relationship between battery cells and equalizers across different equalization structures. 
In more intuitive terms, since different types of equalizers facilitate energy transfer between varying numbers of battery cells, it is difficult to find a general law at the cell level 
accurately describing battery equalization at the pack level. This issue is particularly pronounced in systems like the layer-based CC equalization system, where each cell is connected to many equalizers, making the development of a unified model 
cumbersome and complex.

\subsection{Equalization Current Analysis}
According to Section~\ref{Sec:OverviewA}, the equalizers include CC, MM, CPC, and CMC equalizers, depending on the number of cells they are connected to. 
\subsubsection{Equalization currents through CC equalizers}
To achieve battery equalization, a CC equalizer transfers energy from its connected cell with a high SOC to the other cell with a low SOC. According to the law of conservation of energy \cite{Dorf2016}, for the $l$-th CC equalizer connected with the $i$-th and $j$-th battery cells, it can be calculated that 
\begin{align} \label{11}
    \beta_i V_{B_{i}}(k) I_{ec_{i,l}}(k)+\beta_j V_{B_j}(k) I_{ec_{j,l}}(k)=0
\end{align} 
with
$$\left\{\begin{array}{ll}
\beta_i=1, \beta_j=\alpha_{l}, & \text{if}~SOC_{i}(k) \geq {SOC}_{j}(k)  \\[2pt]
\beta_i=\alpha_{l}, \beta_j=1, &  \text{if}~SOC_{i}(k) < {SOC}_{j}(k)\end{array} \right..$$
 In this study, we define the discharging current as positive and the charging current as negative.
Also, the cells' SOC information is assumed to be available.
Then, based on \eqref{11}, the relationship between the equalization currents of the two associated cells can be deduced as
\begin{align} \label{12}
I_{ec_{i,l}}(k)=- \frac{\beta_j V_{B_j}(k)}{\beta_i V_{B_{i}}(k)} I_{ec_{j,l}}(k).
\end{align}

\begin{ass} \label{A1}
The terminal voltage difference of cells in a battery pack is negligible. 
\end{ass}

\begin{ass} \label{A2}
The energy loss during the cell balancing process caused by equalizers and the associated cable is negligibly small. 
\end{ass}

Based on Assumptions \ref{A1} and \ref{A2}, \eqref{12} can be simplified as 
\begin{align} \label{13}
I_{ec_{i,l}}(k)=-I_{ec_{j,l}}(k).
\end{align}

\begin{rem} 
Assumptions \ref{A1} and \ref{A2} do not hold in practice but are commonly adopted in the model development for battery balancing systems, as seen in \cite{9477598, 6832590}. 
In practice, the cell’s open circuit voltage remains flat in a typical SOC operation range ($20\%–90\%$) \cite{7458184}. A large SOC imbalance is projected on a small difference in the cell voltages. Here, these assumptions are applied to simplify the relationship between the cell equalization currents through one equalizer to uncover the common mathematical characteristics of all the considered equalization structures. These characteristics will not be affected by the energy loss and cell voltage changes, and the generality is maintained. These assumptions can easily be removed by adding the corresponding coefficients to \eqref{13}. 
 \end{rem}

By defining $I_{{ec}_l}$ as the directed equalization current provided by the $l$-th CC equalizer,
it yields 
\begin{subequations}\label{3211}
\begin{align} 
	I_{ec_{i,l}}(k)=\:& I_{ec_l}(k) \label{3211a} \\	
	 I_{ec_{j,l}}(k)=\:&- I_{ec_l}(k).	 \label{3211b} 
\end{align}
\end{subequations}
Referring to \cite{6832590, 8445645, 9328206, 9681274}, $I_{ec_l}$ is commonly defined as
\begin{align}  \label{321133}
I_{ec_l}(k)= \text{sgn}(SOC_i(k)-SOC_j(k)) \bar{I}_{ec_l}(k)
\end{align}
where $\text{sgn}(\cdot)$ is the sign function,  and $\bar{I}_{ec_l}(k)\geq 0$ is the magnitude of the equalization current. 
According to the definition in \eqref{3211}-\eqref{321133}, the equalizer $l$ transfers energy from cell $i$ to cell $j$ when $SOC_i(k)>SOC_j(k)$ and conversely, it transfers energy from cell $j$ to cell $i$ when $SOC_i(k)<SOC_j(k)$. 
When $SOC_i(k)= SOC_j(k)$, there is no balancing operation, and $I_{ec_{i,l}}(k)= I_{ec_{j,l}}(k)=0$.

 Note that for a CC equalizer in the series-based and module-based CC equalization systems, as shown in Fig.~\ref{model}(a--b), we have $j=i+1$ and $l=i$ in \eqref{3211}, since the CC equalizer $i$ is connected to the adjacent cells $i$ and $i+1$. For the first layer in the layer-based CC equalization system, we have $j=i+1$ and $l=\frac{i+1}{2}$, as shown in Fig.~\ref{model}(c). 

\subsubsection{Equalization currents through MM equalizers}
Similar to CC equalizers, an MM equalizer transfers energy from its connected module with a high average SOC to the other module with a low average SOC. Consider the $l$-th MM equalizer connected to the $i$-th battery module containing $b$ series-connected cells labeled $\{c, c+1, \cdots, c+b-1\}$ and the $j$-th battery module containing another $b$ series-connected cells $\{d, d+1, \cdots, d+b-1\}$, where $c$ and $d$ are the cell starting indices in the corresponding modules. Equal voltages and no energy losses imply that the equalization current leaving one module must be received by the other, and because of the series connections, the same current goes through all cells in the module, i.e.,
\begin{subequations}	 \label{323}
\begin{align} 
	I_{em_{c,l}}(k)=\:& I_{em_{c+1,l}}(k)=\cdots=I_{em_{c+b-1,l}}(k) \nonumber
	\\  =\:&I_{em_l}(k) 
 \\	
	I_{em_{d,l}}(k)=\:& I_{em_{d+1,l}}(k)=\cdots=I_{em_{d+b-1,l}}(k) \nonumber
	\\  =\:&-I_{em_l}(k).	 
\end{align}
\end{subequations}
Similar to CC equalizers, $I_{em_l}$ is usually designed as
\begin{align}\label{e11}
I_{em_l}(k)=\text{sgn}(\bar{SOC}_{m_i}(k)-\bar{SOC}_{m_j}(k)) \bar{I}_{em_l}(k)
\end{align}
where $\bar{I}_{em_l}(k) \geq 0$. The average SOCs of the $i$-th and $j$-th battery modules $\bar{SOC}_{m_i}$ and $\bar{SOC}_{m_j}$ are defined by
\begin{align*}
    \bar{SOC}_{m_i}(k)=\:&\frac{1}{b}\sum_{q=c}^{c+b-1} SOC_{q}(k)
    \\
    \bar{SOC}_{m_j}(k)=\:&\frac{1}{b}\sum_{q=d}^{d+b-1} SOC_{q}(k).
\end{align*}
 Note that the MM equalizers are utilized in the module-based CC, layer-based CC, and module-based CPC equalization systems, as illustrated in Fig.~\ref{model}(b-c, e).

\subsubsection{Equalization currents through CPC equalizers}
The CPC equalizer $l$ enables energy transfer 
between its connected cell $i$ and the battery pack
whenever their SOC difference exceeds a given small threshold. The equalization current on the pack side is $\frac{1}{n}$ of the equalization current on the cell side. Moreover, since the $i$-th cell is included in the battery pack, it also receives the same equalization current on the side of the battery pack. Hence, the cell's  equalization current through the $l$-th CPC equalizer $I_{ep_{p,l}}$ ($1 \leq p \leq n$) can be calculated by \cite{9328206}
	\begin{align}\label{35}
			I_{ep_{p,l}}(k)=\begin{cases}
    \frac{n-1}{n} I_{ep_{l}}(k), & p =i\\[2pt]
    -\frac{1}{n} I_{ep_{l}}(k), & p =1, 2, \cdots, n, p \neq i
   \end{cases}
	\end{align}
where $I_{ep_{l}}$ is commonly designed as
\begin{align}\label{e12}
I_{ep_{l}}(k)=\text{sgn}(SOC_{i}(k)-\bar{SOC}_{P}(k)) \bar{I}_{ep_l}(k)
\end{align}
with $\bar{I}_{ep_l} (k) \geq 0$ . The average SOC of the battery pack $\bar{SOC}_{P}$ is defined as 
\begin{align*}
\bar{SOC}_{P}(k)=\frac{1}{n}\sum_{i=1}^{n} SOC_{i}(k).
\end{align*}
Note that a CPC equalizer will generate unequal equalization currents on its connected cell and battery pack. CPC equalizers can be seen in the CPC, module-based  CPC, and switch-based CPC equalization systems, as illustrated in Fig.~\ref{model}(d--f). For the switch-based CPC equalization system, the CPC equalizer transfers energy from the cell with the highest SOC to the battery pack at each sampling step, indicating a variable connection between the cell and pack.

\subsubsection{Equalization currents through CMC equalizers}
The CMC equalizers have the same structure as the CPC equalizers. Their only difference is that one side of the CMC equalizer is connected to the battery module instead of the entire battery pack. For the $l$-th CMC equalizer connected to the $i$-th cell and the $j$-th battery module (containing cells $c, \cdots, i, \cdots, c+b-1$), similar to \eqref{35}, the cell  equalization current through the $l$-th CMC equalizer $I_{ecm_{p,l}}$ ($c \leq p \leq c+b-1$) is 
	\begin{align}\label{353}
			I_{ecm_{p,l}}(k)=\begin{cases}
    \frac{b-1}{b} I_{ecm_{l}}(k) , & p =i\\[2pt]
    -\frac{1}{b} I_{ecm_{l}}(k) , & p =c, \cdots, c+b-1, p \neq i
   \end{cases}
	\end{align}
where $I_{ecm_{l}}$ is usually designed as
\begin{align} \label{e13}
I_{ecm_{l}}(k) = \text{sgn}(SOC_{i}(k)-\bar{SOC}_{m_j}(k)) \bar{I}_{ecm_l}(k)
\end{align}
with $\bar{I}_{ecm_l}(k) \geq 0$.
Note that the CMC equalizers are utilized in the module-based CPC equalization system, as shown in Fig.~\ref{model}(e).

\section{Unified  Model Development for Active Battery Equalization Systems}
The hypergraph is a powerful tool to explore the underlying relationships between objects \cite{6570719}, which has been applied in various fields, such as social relationships \cite{10208205}, computer vision \cite{9264674}, and citation networks \cite{BAI2021107637}.
Distinguished with traditional graphs, an edge in the hypergraph can connect any number of vertices \cite{bretto2013hypergraph}. Since one equalizer may connect more than two battery cells, the hypergraph could be a good candidate to reveal the intrinsic relationship between cells and equalizers by treating them as vertices and edges. This section innovatively introduces hypergraphs to the battery field and judiciously uses the properties of hypergraphs to the modeling of active battery equalization systems.
\subsection{
Hypergraph representation of equalizers and cells
}\label{subsubsection_hypergraph} 
\begin{figure*}[!htb]
	\begin{center}
		\includegraphics[width=0.85\linewidth, trim=10 5 10 5, clip]{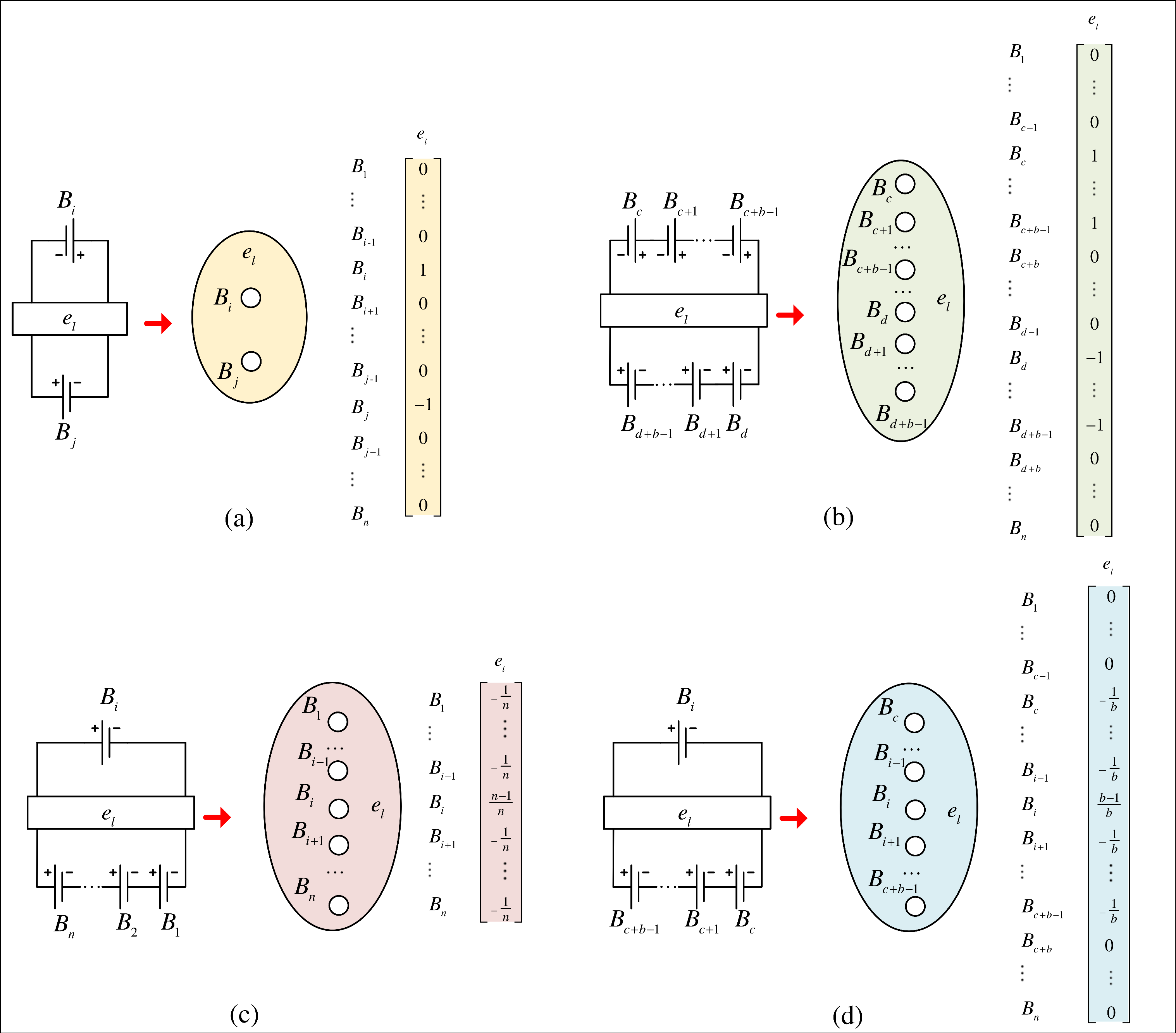}    
		\caption{Electric circuit presentation and equivalent hyperedge of different equalizers: (a) CC, (b) MM, (c) CPC, and (d) CMC equalizers.}
		\label{equalizers}
	\end{center}
\end{figure*}

To uncover the intrinsic connection between battery cells and equalizers, the cells are regarded as the vertices, and the equalizers that transfer energy between cells are treated as edges. Note that the MM, CPC, and CMC equalizers are connected to more than two battery cells, as seen in Fig.~\ref{model}, which means that an edge connects several vertices. These edges are called hyperedges in the concept of hypergraphs \cite{bretto2013hypergraph}. For an active equalization system with $n_e$ equalizers for a battery pack with $n$ series-connected cells, the battery cells are labelled as $B_1, \cdots, B_n$, the CC equalizers are labelled as $e_1,\cdots, e_{n_1}$, the MM equalizers are labelled as $e_{n_1+1}, \cdots, e_{n_2}$, the CPC equalizers are labelled as
$e_{n_2+1}, \cdots, e_{n_3}$, and the CMC equalizers are labelled as $e_{n_3+1}, \cdots, e_{n_e}$, respectively. Then, the active equalization system can be represented by a hypergraph $G=(\nu, \varepsilon)$ with the vertex set $\nu=\{B_1, B_2, \cdots, B_n\}$ and the hyperedge set $\varepsilon =\{e_1, e_2, \cdots, e_{n_e}\}$. To visualize the hypergraphs, in Fig.~\ref{equalizers}, each hyperedge is represented as a big ellipse distinguished by different colors, enclosing all the connected vertices.

The hyperedges $e_l~(1 \leq l \leq n_e)$ are ordered pairs of disjoint subsets of vertices, denoted as $e_l=\{H(e_l), T(e_l)\}$, where $H(e_l)$ and $T(e_l)$ denote the head and tail of $e_l$, respectively \cite{GALLO1993177}. The element of the incidence matrix $C \in$ $ \mathbb{R}^{n \times n_e}$ of the hypergraph $G$, denoted as $c_{p,l}$ $(1 \leq p \leq n, 1 \leq l \leq n_e)$, can then be defined as
\begin{align} \label{3212}
		c_{p,l}=\begin{cases}
			w_{l_h},& \text{if $B_p \in H(e_l)$}\\
			w_{l_t},& \text{if $B_p \in T(e_l)$}\\
			0,&\text{otherwise}
		\end{cases}
\end{align}
where $w_{l_h}$ and $w_{l_t}$ are the weights. 
Note that the tails and heads of the hyperedges constitute two subsets of the battery cells transferring energy through the equalizers.  $c_{p,l} = 0$ implies that the battery cell $p$ has no connection to the equalizer $l$. For the two subsets of battery cells connected with the equalizer $l$, we can randomly select one of them as the head and the other one as the tail. Without loss of generality, we define the top subsets of cells in Fig. \ref{equalizers} as the heads and the bottom subsets as the tails.

The incidence vector corresponding to the hyperedge $e_l$, denoted as $c_l \in \mathbb{R}^{n}$, can be generally formulated as
\begin{align} \label{32123}
c_l=\:& [c_{1,l}, c_{2,l}, \cdots, c_{n,l}]^T \nonumber \\ 
= \:&  [\cdots, 0, \underbrace{w_{l_h}, \cdots, w_{l_h}}_{B_i \in H(e_l)},  0,  \cdots,  0, \underbrace{w_{l_t},\cdots, w_{l_t}}_{B_j \in T(e_l)}, 0, \cdots]^T.
\end{align}
Except for the elements corresponding to vertices within the head or tail subsets, all other elements in the column vector \eqref{32123} are zero. 
The weights $w_{l_h}$ and $w_{l_t}$ can be derived based on the relationship between the equalizing currents on both sides of the equalizer.

\subsection{Application of hypergraphs to equalizer modeling
}\label{sec:hypergraphs2equalizer}
By using the hypergraph theory and its properties described in Section~\ref{subsubsection_hypergraph}, a unified model can be developed for all the active battery equalization systems.

According to the definition of the incidence matrix in the hypergraph theory, the equalization current received by cell $p$ from equalizer $l$ satisfies
\begin{align}
    I_{e_{p,l}}(k) = c_{p,l} I_l(k)\label{eq:element-wise}
\end{align}
where $I_{l}$ represents the directed equalization current provided by equalizer $l$ in any type of active equalization system. For example, $I_{l}$ can be embodied as $I_{{ec}_l}$, $I_{{em}_l}$, $I_{{ep}_l}$, and $I_{{ecm}_l}$. By vectorizing the equalization current received by the in-pack battery cell from equalizer $l$ as $I_{e_l}=[I_{ec_{1,l}}, \cdots, I_{ec_{n,l}}]^T$, based on \eqref{eq:element-wise}, one can readily obtain
\begin{align}
    I_{e_l}(k) = c_l I_l(k). \label{eq:vector-wise}
\end{align}

\subsubsection{Hypergraph-based modeling of CC equalizers}\label{sec:CC-equalizer-model}
For the CC equalizer (hyperedge) $e_l~(1 \leq l \leq n_1)$, as shown in Fig. \ref{equalizers}(a), we have $H(e_l)=\{B_i\}$, and $T(e_l)=\{B_j\}$. As per \eqref{32123}, we can obtain
\begin{align} 
c_l= [\cdots, 0, \underbrace{w_{l_h}}_{p=i}, 0, \cdots, 0, \underbrace{w_{l_t}}_{p=j}, 0, \cdots]^T.
\end{align}
Based on \eqref{3211} and \eqref{eq:element-wise}, we conclude that
\begin{align}
w_{l_h}=1, \: w_{l_t}=-1.
\end{align}
In fact, for any CC equalizer $e_l$ connecting cell $i$ and $j$, 
\begin{align} \label{18}
    I_{e_l}(k) = [\cdots, 0, \underbrace{1}_{p=i}, 0, \cdots, 0, \underbrace{-1}_{p=j}, 0, \cdots]^T I_{{ec}_l}(k) 
\end{align}
where $1 \leq l \leq n_1$. 

\subsubsection{Hypergraph-based models of other equalizers}
For the MM equalizer $e_l~(n_1+1 \leq l \leq n_2)$ illustrated in Fig.~\ref{equalizers}(b), $H(e_l)=\{B_c,B_{c+1}, \cdots, B_{c+b-1}\}$ and $T(e_l) =\{B_d, B_{d+1}, \cdots, B_{d+b-1}\}$. For the  CPC equalizer $e_l~(n_2+1 \leq l \leq n_3)$ Fig.~\ref{equalizers}(c), $H(e_l)=\{B_i\}$, $T(e_l)=\{B_1, \cdots, $ $B_{i-1}, B_{i+1}, \cdots, $ $B_{n}\}$. For the CMC equalizer $e_l~(n_3+1 \leq l \leq n_e)$ illustrated in Fig.~\ref{equalizers}(d), $H(e_l)=\{B_i\}$,  $T(e_l)=\{B_c, \cdots,  B_{i-1}, B_{i+1}, \cdots, B_{c+b-1}\}$. By using the same procedure as in Section \ref{sec:CC-equalizer-model}, we can obtain
\begin{align}
    I_{e_l}(k) =  [ \:& \cdots, 0, \underbrace{1,  \dots, 1,}_{p=c, \cdots, c+b-1} 0 \dots, 0, \underbrace{-1, \dots, -1}_{p=d, \cdots, d+b-1}, \nonumber \\ \:& 0, \cdots]^T   
   I_{{em}_l}(k), ~\: n_1+1 \leq l \leq n_2
  \\ 
    I_{e_l}(k) = \:& [\underbrace{-\frac{1}{n}, \cdots, -\frac{1}{n}}_{p=1,\cdots, i-1}, \underbrace{\frac{n-1}{n}}_{p=i}, \underbrace{-\frac{1}{n}, \dots, -\frac{1}{n}}_{p=i+1,\cdots, n}]^T\nonumber \\  \:& \times I_{{ep}_l}(k), ~\: n_2+1 \leq l \leq n_3 
 \end{align}
\begin{align} \label{20} 
    I_{e_l}(k) = [ \:&\cdots, 0,  \underbrace{-\frac{1}{b}, \cdots, -\frac{1}{b}}_{p=c,\cdots, i-1}, \underbrace{\frac{b-1}{b}}_{p=i}, \underbrace{-\frac{1}{b}, \dots, -\frac{1}{b}}_{p=i+1,\cdots, c+b-1}, \nonumber \\ \:& 0, \cdots]^T I_{{ecm}_l}(k), ~\: n_3+1 \leq l \leq n_e
\end{align}

\begin{figure*}[hbt]
	\begin{center}
		\includegraphics[width=0.98\linewidth]{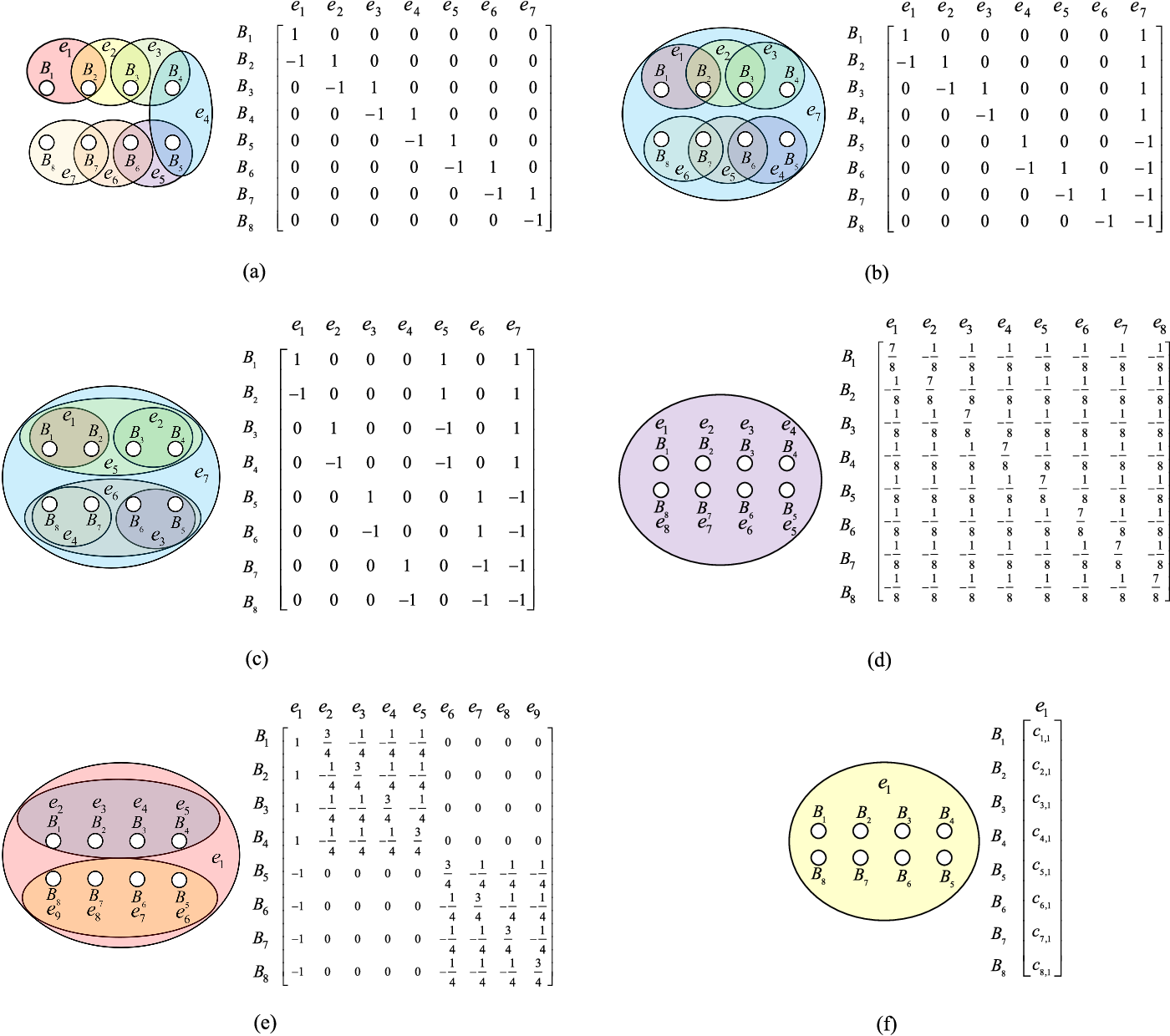}    
		\caption{Hypergraphs and incidence matrices of (a) Series-based CC, (b) module-based CC, (c) layer-based CC, (d) CPC, (e) module-based CPC, and (f) switch-based CPC equalization systems for a battery pack containing $8$ series-connected cells.}
		\label{combined6}
	\end{center}
\end{figure*}

\subsection{
Hypergraph-based battery system modeling
} \label{sec:hypergraphs3}
According to Coulomb counting, the dynamics of the $i$-th ($1 \leq i \leq n$) cell's SOC in the battery pack can be described by \cite{8888246}:
%
\begin{align} \label{32344}
		SOC_i(k+1)= SOC_i(k)-\frac{\eta T_0}{3600 Q_i} (I_s(k)+I_{eq_i}(k))
\end{align}
where the coulombic efficiency $\eta$ can here be assumed to be $1$ \cite{9328206}.
According to Kirchhoff's circuit law, the cells' total equalizing currents are equal to the sum of the currents contributed by their connected equalizers. Based on the equalization currents in \eqref{18}--\eqref{20}, the total equalization current vector for the cells in the battery pack $I_{eq} = [I_{eq_1}, I_{eq_2}, \cdots, I_{eq_n}]^T$ can be calculated as
\begin{align} \label{1111}
I_{eq}(k)=\:& \sum\limits_{l=1}^{n_e} I_{e_l}(k) \nonumber \\
=\:& \sum\limits_{l=1}^{n_1} c_l I_{ec_l}(k)+\sum\limits_{l=n_1+1}^{n_2} c_l I_{em_l}(k) \nonumber  \\
  &  +\sum\limits_{l=n_2+1}^{n_3} c_l I_{ep_{l}}(k)  +\sum\limits_{l=n_3+1}^{n_e} c_l  I_{ecm_{l}}(k).
\end{align}
Then, by defining the state vector $x(k)$, the diagonal matrix $D$, the vector $d(k)$, and the control variable vector $u(k)$ as 
\begin{align*}
  x(k)= \:& 
  [SOC_1(k), SOC_2(k), \cdots, SOC_n(k)]^T \in \mathbb{R}^{n}  \nonumber \\
  D = \:& \text{diag}\left\{\frac{\eta T_0}{3600 Q_1}, \cdots, \frac{\eta T_0}{3600 Q_n}\right\} \in \mathbb{R}^{n \times n} \\
  d(k) 
  = \:& [I_s(k), \cdots,I_s(k)]^T  \in \mathbb{R}^{n} 
   \\
  u(k) = \:& \left[I_{ec_1}(k),\cdots,  I_{ec_{n_1}}(k), I_{em_{n_1+1}}(k), \right. \\
  & \:\: \cdots,   I_{em_{n_2}}(k), I_{ep_{n_2+1}}(k), \cdots, I_{ep_{n_3}}(k), 
  \\
  & \left. I_{ecm_{n_3+1}}(k), \cdots, I_{ecm_{n_e}}(k) \right] \in \mathbb{R}^{n_e} 
\end{align*}
where $\text{diag}(\cdot)$ represents a diagonal matrix, the unified model for battery equalization systems can be formulated 
in the following state-space form:
\begin{align} \label{321}
		x(k+1)=\:&x(k)-D  C u(k) -D d(k). 
\end{align}
where 
$C=[c_1, \cdots, c_{n_e}]$ is the incidence matrix. 

From  (\ref{321}), it is observed that the key to obtaining the model of battery equalization systems is to determine the incidence matrix $C$ of its equivalent hypergraph. 
To illustrate, we consider a pack with $n=8$ series connected cells in $m=2$ modules.
The equivalent hypergraphs and incidence matrices of the previously mentioned six battery equalization systems are shown in Fig. \ref{combined6}. It should be pointed out that for the switch-based CPC equalization system, $C$ is a variable matrix depending on which cell has the largest SOC. Specifically, $c_{i,1}=\frac{n-1}{n}$ if $B_i$ has the largest SOC among all the considered cells, otherwise $c_{i,1}=-\frac{1}{n}$.

Note that the developed unified model (\ref{321}) is not limited to the six battery equalization systems studied in this work but can be easily extended to any equalization system with the commonly used CC, MM, CPC, and/or CMC equalizers.

\section{Equalization Performance Analysis based on the Unified Model
}
\subsection{The Minimum Required Number of Equalizers} 
A controllability analysis will be used to determine whether the battery equalization systems can regulate the SOC of all cells to the same level. 
A manifold $\Gamma$ is defined as where the cells' SOCs in the battery pack are identical, i.e., \cite{Yang14}: 
\begin{equation} \label{3231}
	\begin{array}{lll}
		\Gamma \triangleq \{x  \in \mathbb{R}^{n}| x_1=x_2=\cdots=x_n\}.
	\end{array}
\end{equation}
To satisfy (\ref{3231}), only $n-1$ states among these $n$ states need to be controlled. For example, if we can make $\{x_2-x_1=0, x_3-x_1=0,\cdots, x_{n}-x_1=0\}$, the SOCs of all the $n$ cells are identical. With this in mind, we can define a new $n-1$ dimensional state variable $s$ for controllability analysis of the battery system, given by
\begin{align} \label{B111}
    s(k)= L x(k)
\end{align}
where $s\in\mathbb{R}^{n-1}$, and 
$L \in \mathbb{R}^{(n-1) \times n}$. 
For the exemplified case of $\Gamma$, $L$ becomes
\begin{align*}
		L=\begin{bmatrix}
			-1	& 1 & 0 & \cdots & 0 & 0\\ 
			-1	& 0 & 1 & \cdots & 0 & 0\\
			\vdots 	& \vdots & \vdots & \ddots & \vdots & \vdots \\
			-1	& 0  & 0 & \cdots & 0 & 1
		\end{bmatrix}_{(n-1) \times n}
\end{align*}
and correspondingly, $s(k)=0_{n-1}$ if and only if $x(k) \in \Gamma$, 
where $0_{n-1}$ denotes the column zero-vector in $n-1$ dimensions. 
Inserting (\ref{B111}) into (\ref{321}),
\begin{align} \label{32331}
		s(k+1)=s(k)- LDCu(k)-LDd(k),
\end{align}
which is in a standard linear state-space form.

\begin{lemma} \label{L1}
For any battery system composed of $n$ series-connected cells with the state dynamics governed by \eqref{321}, given a fixed incidence matrix $C$, defined in \eqref{3212}, at least $n-1$ equalizers are required to achieve equalization of all the $n$ battery cells.
\end{lemma}

\begin{IEEEproof}
Based on control theory \cite{ogata2010modern}, the system (\ref{32331}) is controllable, i.e., the state can be transferred from any initial state $s(0)=Lx(0)$ to the final state $s(k)=0_{n-1}$ in finite time by a control sequence  $\{u(k)\}$, 
if the following condition is satisfied:
\begin{align} \label{323321}
		\text{rank}([LDC, ILDC, \cdots, I^{n-1}LDC]) 
		=\text{rank}(LDC)=n-1
\end{align} 
where the identity matrix $I$ is the state-transition matrix in (\ref{32331}), and $\text{rank}(\cdot)$ outputs the rank of a matrix. 
Since the system's transition matrix is invertible, this condition is both sufficient and necessary \cite{K97}.
Since $\text{rank}(L)=n-1$ and $\text{rank}(D)=n$, it can be obtained that
\begin{align} \label{33321}
		\text{rank}(LDC) \leq \text{min}(n-1,\text{rank}(C)).
\end{align}
From (\ref{323321}) and (\ref{33321}), the necessary condition for the battery equalization systems, i.e., \eqref{321} with a fixed $C$, to be controllable can be derived
\begin{align}  \label{3321}
		\text{rank}(C) \geq n-1.
\end{align} 
Given the number of equalizers equals to the number of columns in the incidence matrix $C$, (\ref{3321}) means that the considered equalization systems need at least $n-1$ equalizers. This completes the proof. 
\end{IEEEproof}
%

For the series-based CC, module-based CC, layer-based CC, CPC, and module-based CPC equalization systems, their corresponding ranks of $C$ are all $\text{rank}(C)=n-1$. Therefore, all these equalization systems are controllable. If we reduce one equalizer in the series-based CC/module-based CC/layer-based CC equalization system, the number of equalizers decreases to $n-2$, and the rank of $C$ is reduced to $\text{rank}(C)=n-2$, which produces an uncontrollable system (\ref{32331}). For the CPC equalization system, there are $n$ equalizers. If we remove one equalizer, since its rank of $C$ is reduced to $\text{rank}(C)=n-1$, this equalization system is still controllable, indicating that there is one redundant equalizer. Only $n-1$ CPC equalizers can achieve the equalization of the battery pack with $n$ series-connected cells. Similarly, the module-based CPC equalization system is still controllable if the MM equalizers and one CPC equalizer are removed.
 
 The battery equalization system with a variable $C$, such as the switch-based CPC system, is capable of attaining the same results as the standard CPC equalization system, as it also allows for energy to be transferred between any cell and the battery pack. 
The switch-based equalization system requires a longer time to achieve balance, because only one, or a small number of equalizers, are active at any given moment.
 
\subsection{Comparison of Different Structures in Equalization time}
By using the developed unified model (\ref{321}), we can conveniently estimate and compare the equalization time of various battery equalization systems. 

The equalization time $T_e$ can be defined as the minimum time for the SOCs of all the in-pack battery cells to converge to the vicinity of the manifold $\Gamma$, i.e., 
\begin{align}  \label{333231}
		T_e = \text{min}\{\tau |~ \frac{1}{n}||x(k)-\bar{x}(k)|| \leq \epsilon, \forall k T_0 \geq \tau \}
\end{align} 
where $||\cdot||$ stands for the 2-norm, $\bar{x}(k)=\frac{1}{n} 1_n 1_n^T x(k)$ with $1_n$ being the column vector with $n$ ones. Below, Algorithm~\ref{algo_AA} is proposed to estimate the equalization time.

\begin{algorithm} 
\caption{}\label{algo_AA} 
\begin{enumerate}
\item Set the cells' capacity $Q_i$ and initial state-of-charge, $SOC_i(0)$ ($1 \leq i \leq n$), the tolerance bound $\epsilon$, and the sampling period $T_0$.

\item  Based on the connection of CC/MM/CPC/CMC equalizers (hyperedges), set the vectors $c_l$ $(1 \leq l \leq n_e)$ through (\ref{32123}) to obtain the incidence matrix $C=[c_1, \cdots, c_{n_e}]$. 

\item 
If there is a switching network in the equalization system, jump to Step 5); otherwise, continue to Step 4).

\item  Terminate and output ``The equalization cannot be achieved!", if (\ref{323321}) is not satisfied. Otherwise, continue to Step 5).

\item Iterate $x(k)$ based on the state-space model (\ref{321}) while ignoring the external current, i.e., setting $d(k)=0_n$. 

\item Terminate and output $T_e=kT_0$, if $\frac{1}{n}||x(k)-\bar{x}(k)||\leq \epsilon$. Otherwise, set $k=k+1$ and return to Step 2).
\end{enumerate}
\end{algorithm}


\begin{lemma} \label{L2}
The equalization time is upper bounded according to
\begin{equation}\label{TeBound}
	T_e \leq \frac{\text{log}(||(x(0)-\bar{x}(0))||)-\text{log}(n \epsilon)}{-\text{log}(1-d_sk_s\lambda_{n-1}(CC^T))}T_0,
\end{equation}
where $d_s=\eta T_0/(3600\max{Q_i})$, $k_s$ is the smallest feedback gain of all the SOC differences, and $\lambda_{n-1}$ denotes the second smallest eigenvalue.
\end{lemma}

\begin{IEEEproof}
By substituting the control laws \eqref{321133}, \eqref{e11}, \eqref{e12}, and \eqref{e13} into \eqref{321}, it can be obtained that
\begin{align} 
			x(k+1)=\:&x(k)-D C  \left[\text{sgn}(C^T x(k))\circ v(k)\right] -D d(k)   \label{321sss}
\end{align}
where $\circ$ is the Hadamard product \cite{horn2012matrix}, and $v$ is the vector of all the equalization currents in magnitude, defined as
\begin{align}
v = \:& \left[\bar{I}_{ec_1}, \cdots, \bar{I}_{ec_{n_1}}, \bar{I}_{em_{n_1+1}}, \cdots,  \bar{I}_{em_{n_2}}, \right. \nonumber \\
& \:\: \left. \bar{I}_{ep_{n_2+1}}, \cdots,
\bar{I}_{ep_{n_3}},  \bar{I}_{ecm_{n_3+1}}, \cdots, \bar{I}_{ecm_{n_e}}\right]^T. \nonumber
\end{align}
Suppose $v(k)$ has the form $v(k)=K(k)|C^Tx(k)|$, where $|C^Tx(k)|$ is a vector with elements taking the absolute value of the elements of $C^Tx(k)$, and $K(k)$ is a positive diagonal matrix (considering the positivity of $v$). Note that this control protocol is a general form since $K(k)$ can be selected as $K(k)=\phi(x(k))$ with $\phi(\cdot)$ being any positive function. Based on the defined form for $v$, \eqref{321sss} can be re-written as
\begin{align} 
	x(k+1)=\:&x(k)-D C  K(k) C^T x(k) -D d(k)\label{325}
\end{align}
where $K(k) C^T x(k)$ has replaced $\text{sgn}(C^T x(k))\circ v(k)$ in \eqref{321sss} to simplify the model representation.

 By ignoring the external current $d$ for simplification and defining $A(k-1)=I-D C K(k-1) C^T$, the closed-loop system can be deduced from (\ref{325}) as
 \begin{align}  \label{3213}
 		x(k)=A(k-1) x(k-1).
 \end{align}

As per the hypergraph-based equalizer modeling in Section~\ref{sec:hypergraphs2equalizer}, the sum of each column of the incidence matrix $C$ for each type of equalization system is $0$. 
This means that $DCKC^T$ does not have full rank and must have one eigenvalue in 0. Now, denoting the diagonal elements of $D$ by $d_i$ ($i=1,\ldots n$) and of $K$ by $k_j$ ($j=1,\ldots n_e$), the element in row $i$ and column $l$ of $DCKC^T$ is 
\begin{align}
    [DCKC^T]_{il}=\sum_{j=1}^{n_e}d_ic_{ij}k_jc_{lj},
\end{align}
where $c_{ij}$ is the element in row $i$ and column $j$ of $C$. Consequently, the sum of the elements in row $i$ is
\begin{align}
    \sum_{l=1}^n\sum_{j=1}^{n_e}d_ic_{ij}k_jc_{lj}=d_i\sum_{j=1}^{n_e}c_{ij}k_j\sum_{l=1}^nc_{lj}=0
\end{align}
since the column sums of $C$ are zero. From the definition of eigenvector, it then follows that the eigenvector corresponding to the zero eigenvalue of $DCKC^T$ must be $1_n$ because of the zero row sums. Thus, we have $A(k-1)1_n=1_n, \forall k$. Using this, and $\bar{x}(k)=\frac{1}{n} 1_n 1_n^T x(k)$,  while recursively applying \eqref{3213} yields
%
%
%
%
 \begin{align}  \label{32133}
		x(k)-\bar{x}(k)=\:&(A(k-1)-\frac{1}{n} 1_n 1_n^T) \times \cdots \nonumber \\	
		&\times (A(0)-\frac{1}{n} 1_n 1_n^T)(x(0)-\bar{x}(0)).
\end{align} 
Referring to \cite{7995102}, it can then be obtained that
 \begin{align} \label{321323}
		||x(k)-\bar{x}(k)|| 
  \leq ||(A_s-\frac{1}{n} 1_n 1_n^T)||^k ||(x(0)-\bar{x}(0))||
\end{align}
with
$$A_s=I- d_s k_s CC^T,$$
where $d_s$ and $k_s$ are the smallest diagonal elements of $D$ and $K(k-1)$ $(k=0,1,\cdots)$, respectively. 

$CC^T$ is symmetric and positive semi-definite and therefore the eigenvalues $\lambda_i(C C^T)$ are real and non-negative. Further, the sum of the eigenvalues equals the trace of $CC^T$, which is bounded and grows linearly with $n$. Thus, even for the largest eigenvalue, there is a sampling time $T_0$ and gain $K$ such that $d_s k_s \lambda(C C^T)$ is non-negative and much less than~$1$. In fact, since  $d_s$ in practice is very small, so is this product for all reasonable sampling times. As a consequence $0 \leq \lambda_i(A_s) \leq 1,\; \forall i$, where $\lambda_i(\cdot)$ is the $i$-th largest eigenvalue, and the spectral radius of $A_s-\frac{1}{n} 1_n 1_n^T$ can therefore be expressed as \cite{xiao04}
\begin{align} \label{45}
\rho(A_s-\frac{1}{n} 1_n 1_n^T)=1-d_s k_s \lambda_{n-1}(CC^T).
\end{align}
Since $A_s$ is symmetric, the spectral radius equals the induced 2-norm, i.e., 
\begin{align} \label{466}
||A_s-\frac{1}{n} 1_n 1_n^T || = \rho(A_s-\frac{1}{n} 1_n 1_n^T).
\end{align}

Applying the upper bound of $||(x(k)-\bar{x}(k))||$ obtained in~(\ref{321323}) to the definition of equalization time~(\ref{333231}), we have that the smallest integer $k$ satisfying

\begin{align}
   ||(A_s-\frac{1}{n} 1_n 1_n^T)||^k ||(x(0)-\bar{x}(0))|| \leq n \epsilon \label{eq:relaxization}
\end{align}
gives an upper bound of the equalization time. Using \eqref{45} and \eqref{466} and taking the logarithm of \eqref{eq:relaxization} then yields \eqref{TeBound}.
\end{IEEEproof}


\begin{rem}
Any conservatism in the upper bound of $T_e$ originates from the inequality~\eqref{321323}, which in turn has two causes. The first comes from the imbalances in the initial state, i.e. $x(0)-\bar{x}(0)$, and is simply an effect of that the norm of a projection depends on the direction of the input, i.e. the initial state vector. The second cause is the norm approximation of the state transitions by $A_s-\frac{1}{n}1_n1_n^T$. Normally this should not cause any major conservatism. Assuming all $Q_i$ to be equal, and the controller gains $k_i$ to be constants, it follows that $A$ is constant, $A_s=A$, and 
\begin{equation}\label{eq:AsA}
x(k)-\bar{x}(k)=(A_s-\frac{1}{n}1_n1_n^T)^k(x(0)-\bar{x}(0)).
\end{equation}
Since $A$ is real and symmetric so is $(A_s-\frac{1}{n}1_n1_n^T)$, which can then be diagonalized with an orthonormal matrix $U$ to equal $U\Lambda U^T$. Using this in \eqref{eq:AsA} we have that $||(A_s-\frac{1}{n}1_n1_n^T)^k||=||\Lambda ||^k$ without any conservatism, except for the dependence on the initial SOC.
 \end{rem}

 \begin{rem}
From \eqref{TeBound} it is observed that the equalization time is related to the cells' initial SOC distribution $||(x(0)-\bar{x}(0))||$, the number of cells  $n$, the selected tolerance $\epsilon$, the designed control gains $K(k)$, and the incidence matrix of the equalization system $C$. Here, we only consider the effects of the structure of equalization systems, namely the matrix $C$. Since $d_sk_s\lambda_{n-1}(CC^T)$ is small, we may apply a Maclaurin expansion to the denominator of (\ref{TeBound}) to arrive at
\begin{align}\label{43}
		T_e \propto \frac{1}{\lambda_{n-1}(CC^T)},
\end{align}
from which it can be concluded that the smaller $\lambda_{n-1}(CC^T)$ is, the longer the expected equalization time.
\end{rem}

\begin{rem}
   The existing literature, such as recent review articles \cite{9681274, 6832590}, suggests that current methods for comparing different equalization systems depend heavily on extensive and labor-intensive experiments or simulations. These simulations are often based on models specific to each equalization system, requiring sophisticated estimation algorithms tailored to each model. However, with the introduction of Algorithm~\ref{algo_AA}, this challenge can be addressed more efficiently. This approach involves simulating the proposed unified model, \eqref{321}, with various initializations, simplifying the process significantly. 
\end{rem}
\begin{rem} 
Lemma~\eqref{L2} introduces a potentially more efficient method than Algorithm~\ref{algo_AA}. Instead of conducting numerous simulations, it may be sufficient to compare only the second smallest eigenvalues of their corresponding Laplacian matrices $CC^T$. These eigenvalues can be readily obtained through hypergraph-based equalizer modeling, further streamlining the comparison process.
\end{rem}

\section{Results and Discussion}
\subsection{Model Validation}
\begin{figure}[!htb]
	\begin{center}
		\includegraphics[width=0.9\linewidth]{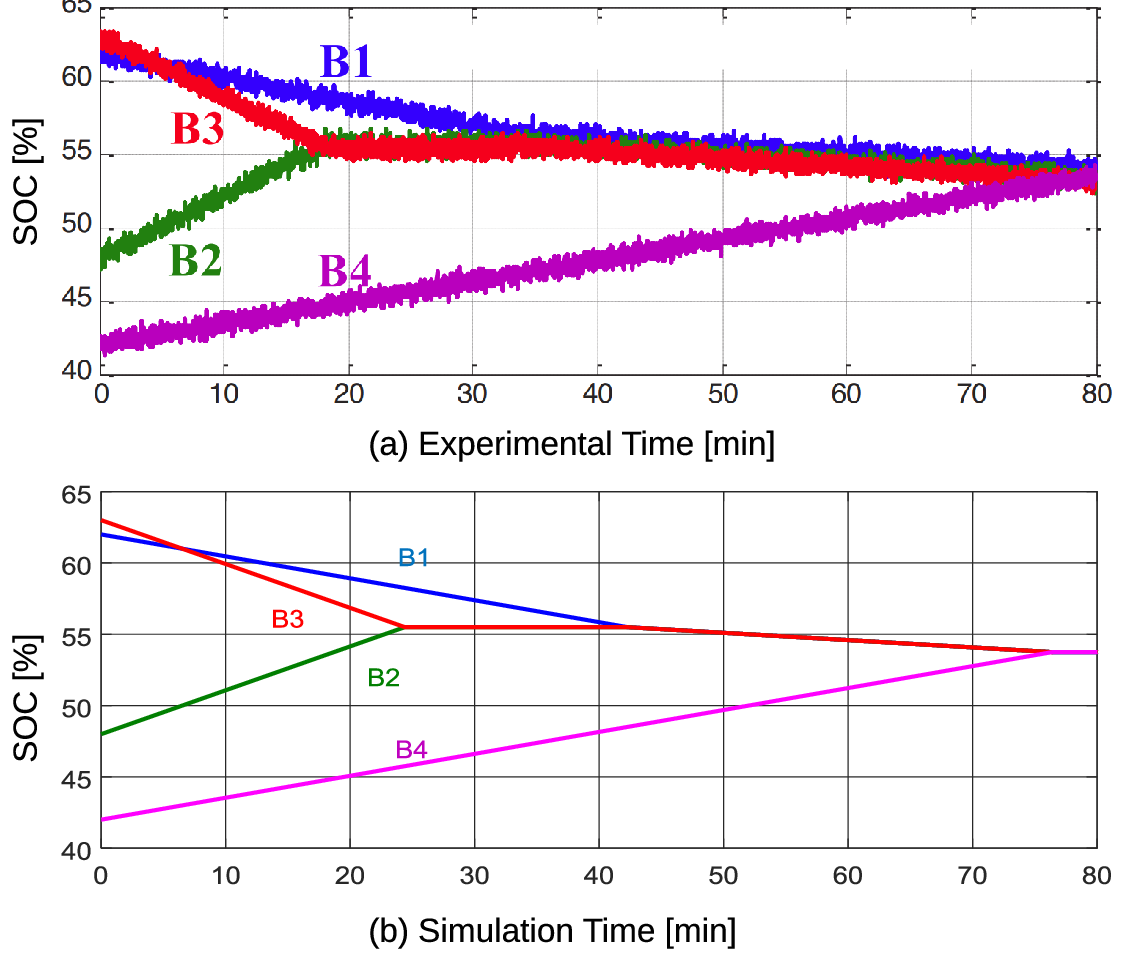}    
		\caption{Model validation results for a series-based CC equalization system. (a) The experimental result obtained in \cite{6063871}. (b) The simulation result from our proposed model (\ref{321}).}
		\label{SeriesC}
	\end{center}
\end{figure}

\begin{figure}[!htb]
	\begin{center}
		\includegraphics[width=0.9\linewidth]{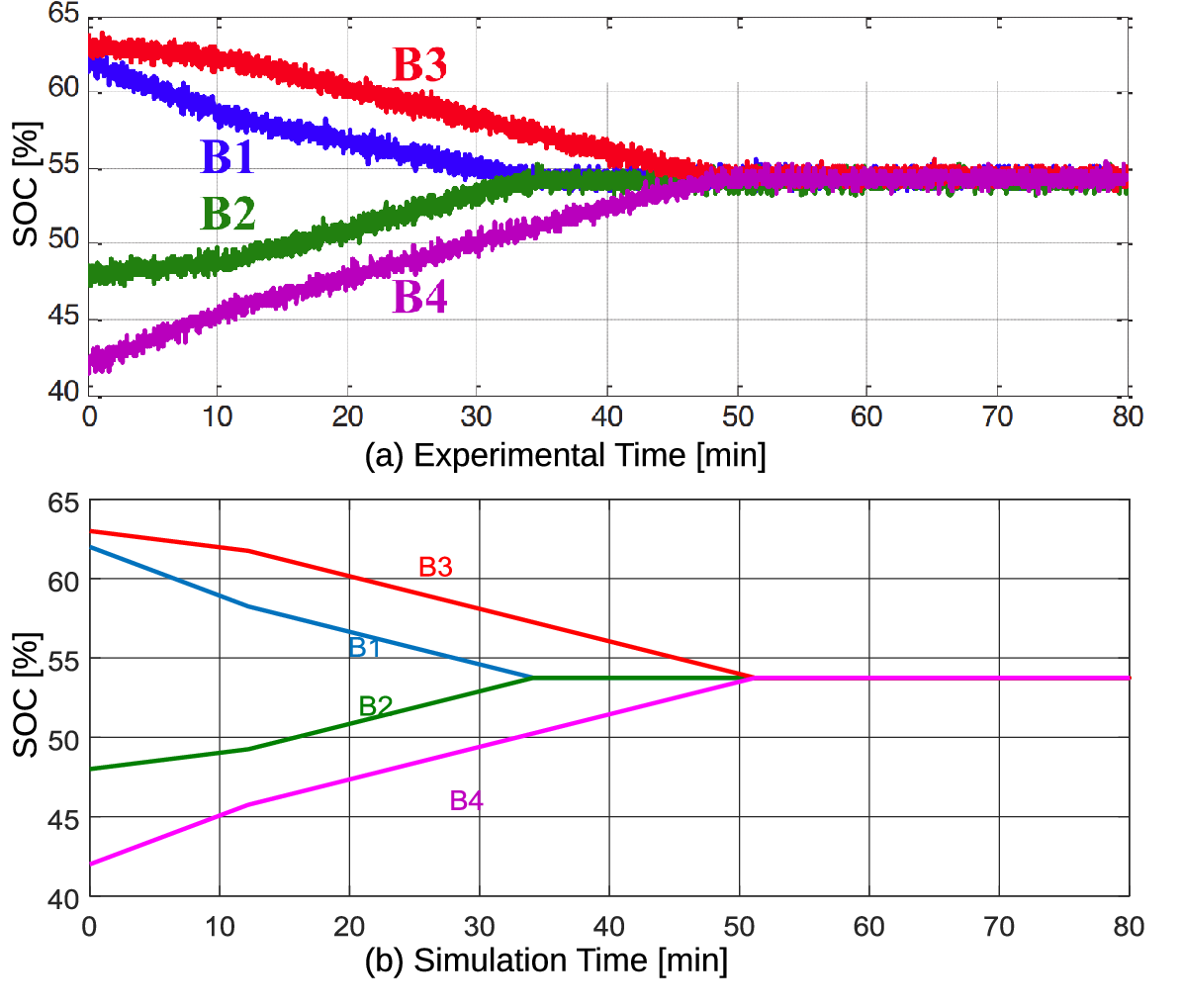}    
		\caption{Model validation results for a layer-based CC equalization system. (a) The experimental result obtained in \cite{6063871}. (b) The simulation result from our proposed model (\ref{321}).}
		\label{Layer}
	\end{center}
\end{figure}

To evaluate the developed unified model \eqref{321}, we first look at its capability of describing the series-based CC equalization system. The experimental results are taken from \cite{6063871}, in which four battery cells with a capacity of $65\rm{Ah}$ were used to generate the SOC evolution trajectories. With the initial SOCs as $x(0)=[62\%, 48\%, 63\%, 42\%]^T$, comparative results between the model-based outputs and measured outputs are presented in Fig.~\ref{SeriesC}, where the experimental SOC was calculated from Coulomb counting, and the fluctuations were due to noise of the utilized current sensors.
It can observed that the model-predicted trajectories are similar to the measured ones. Furthermore, the equalization time is $76.4~\rm{min}$ derived from (\ref{321}), which is close to $80~\rm{min}$ in the experimental result. In addition to series-based CC equalization systems, the proposed model is further assessed on the layer-based CC equalization system, with the results illustrated in Fig.~\ref{Layer}.  Again, a good level of consistency is achieved between the model and measurements, which verifies the developed model.


%

\begin{figure}[!htbp]
	\centering		
	\subfigure[]{
		\begin{minipage} {\linewidth}\label{MCCr}
			\centering
			\includegraphics[width=0.9\linewidth, trim=1 1 20 10, clip]{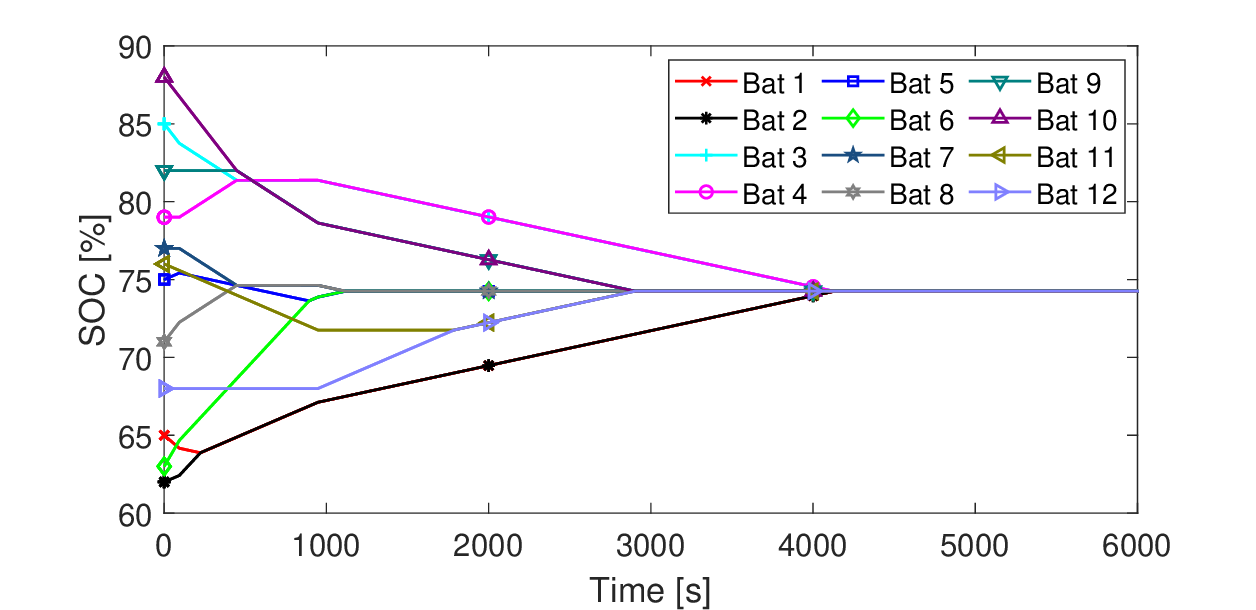}
		\end{minipage}
	}	
	\subfigure[]{
		\begin{minipage}{\linewidth}\label{CPCr}
			\centering
			\includegraphics[width=0.9\linewidth, trim=1 1 20 10, clip]{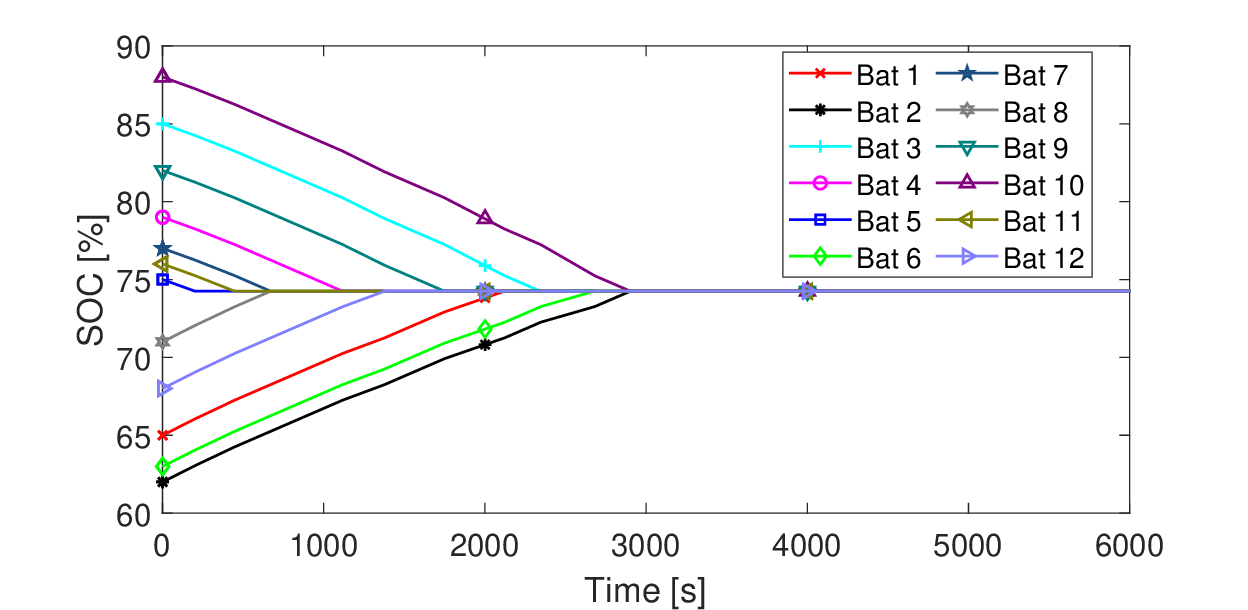}
		\end{minipage}
	}	
	\subfigure[]{
		\begin{minipage} {\linewidth}\label{MCPCr}
			\centering
			\includegraphics[width=0.9\linewidth, trim=1 1 20 10, clip]{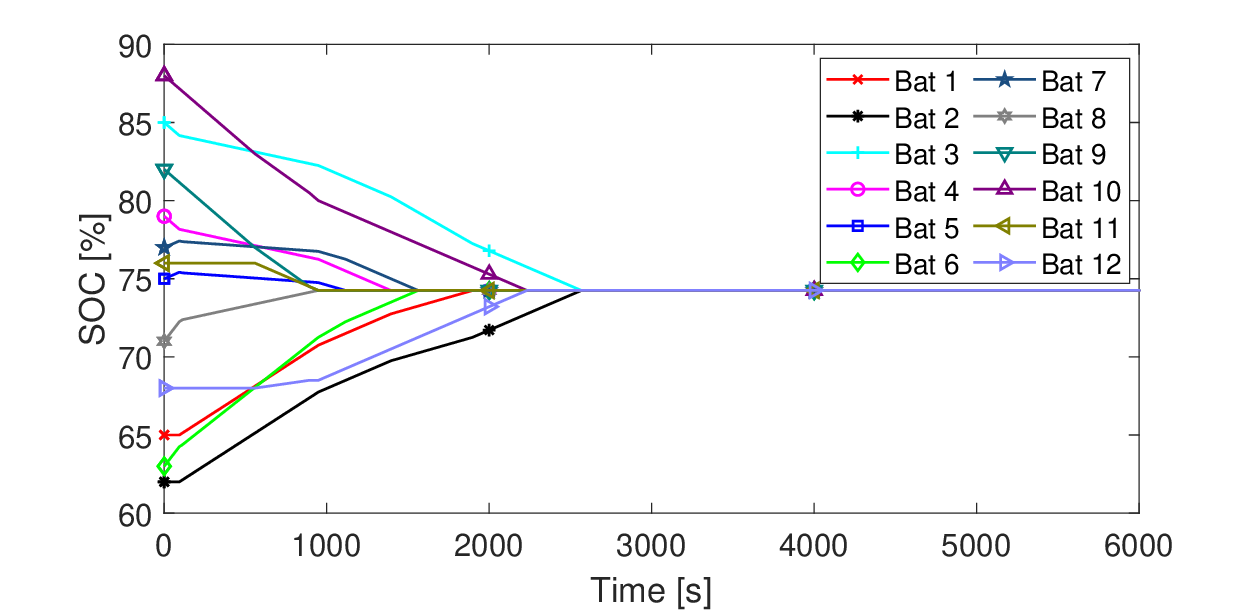}
		\end{minipage}
	}	
	\subfigure[]{
		\begin{minipage} {\linewidth}\label{SCPCr}
			\centering
			\includegraphics[width=0.9\linewidth, trim=1 1 20 10, clip]{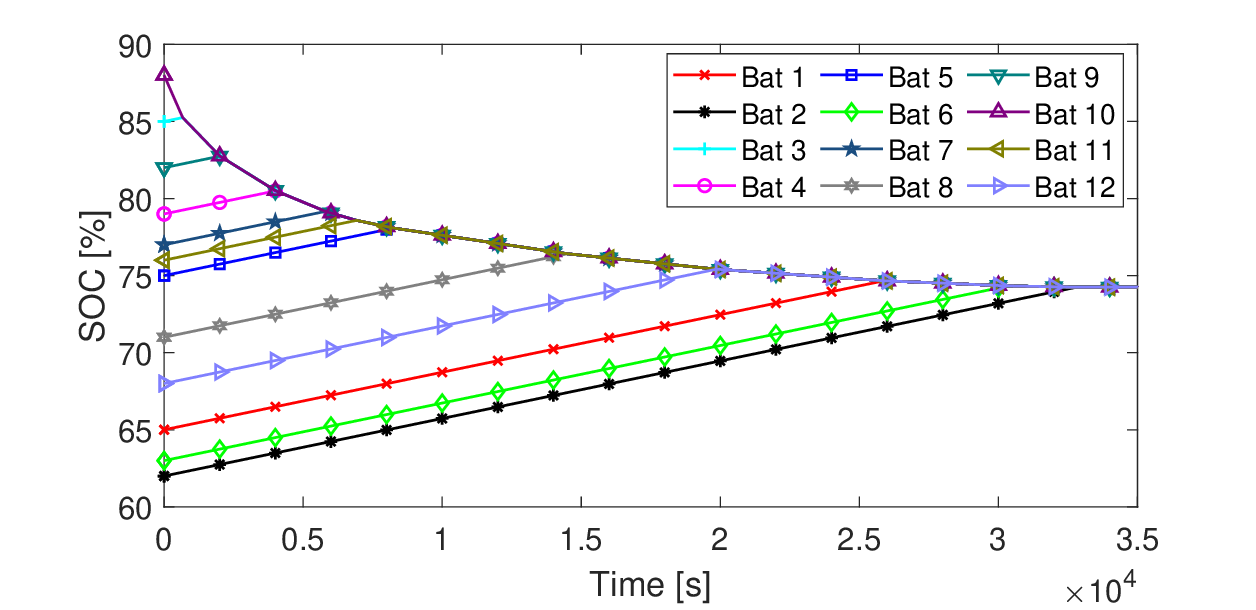}
		\end{minipage}
	}	
	\caption{Simulation results for (a) module-based CC, (b) CPC, (c) module-based CPC, (d) switch-based CPC equalization systems based on model (\ref{321}).}
	\label{Re}
\end{figure}

To further test the proposed battery equalization model on other system structures, a battery pack consisting of $12$ cells with the capacity of $3.1\rm{Ah}$  is utilized, where the initial SOCs of battery cells are set as $x(0)=[65\%,62\%,  85\%, 79\%, 75\%, 63\%, 77\%, 71\%, 82\%, 88\%, 76\%,$ $ 68\%]^T$. The model-based simulation results for the module-based CC, CPC, and module-based CPC systems are illustrated in Figs.~\ref{Re}(a-c), respectively. These results align with those reported in \cite{9328206}, further validating the efficacy of our developed hypergraph-based model. The simulation result for the switch-based CPC equalization system is also given in Fig. \ref{Re}(d).

In the simulations above, we have ignored cell-to-cell differences in terminal voltage and energy transfer losses during cell balancing, following Assumptions~\ref{A1}--\ref{A2}.  Simulation results in Fig. \ref{SeriesC}--\ref{Re} indicate that these assumptions have a limited impact on the convergence of SOC trajectories and the minimum equalization time. 


	\begin{figure}[!htbp]
	\centering		
	\subfigure[]{
		\begin{minipage} {\linewidth}\label{CPCtr}
			\centering
			\includegraphics[width=0.9\linewidth, trim=1 1 20 10, clip]{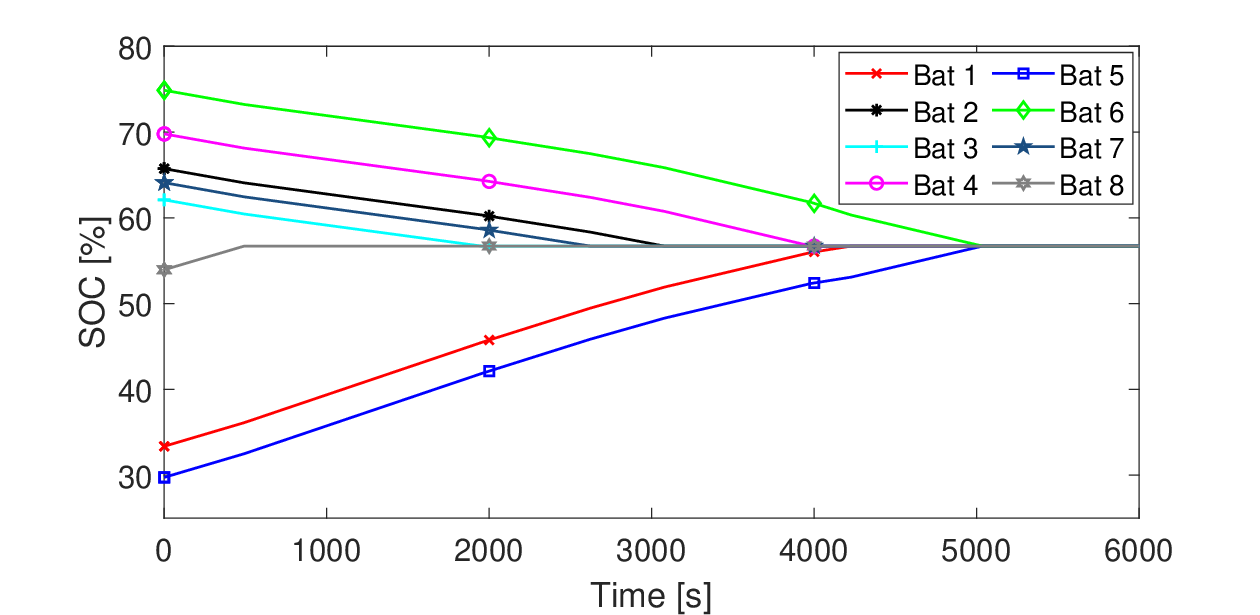}
		\end{minipage}
	}	
	\subfigure[]{
		\begin{minipage}{\linewidth}\label{CPCr2}
			\centering
			\includegraphics[width=0.9\linewidth, trim=1 1 20 10, clip]{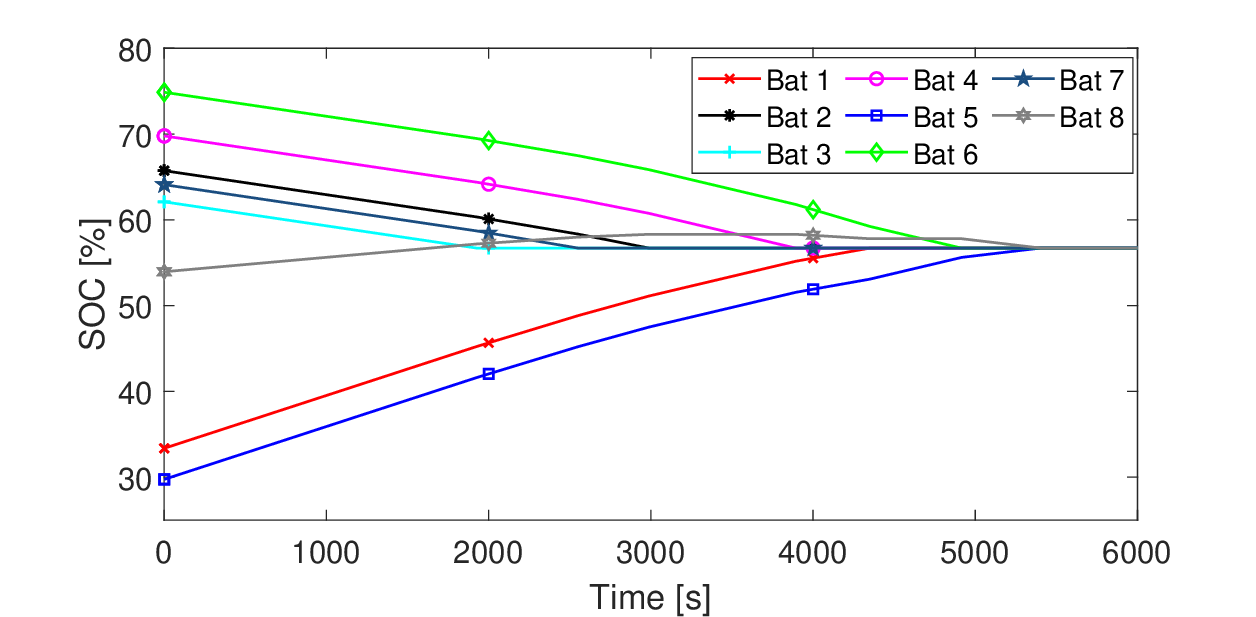}
		\end{minipage}
	}	
	\subfigure[]{
		\begin{minipage} {\linewidth}\label{CPCr3}
			\centering
			\includegraphics[width=0.9\linewidth, trim=1 1 20 10, clip]{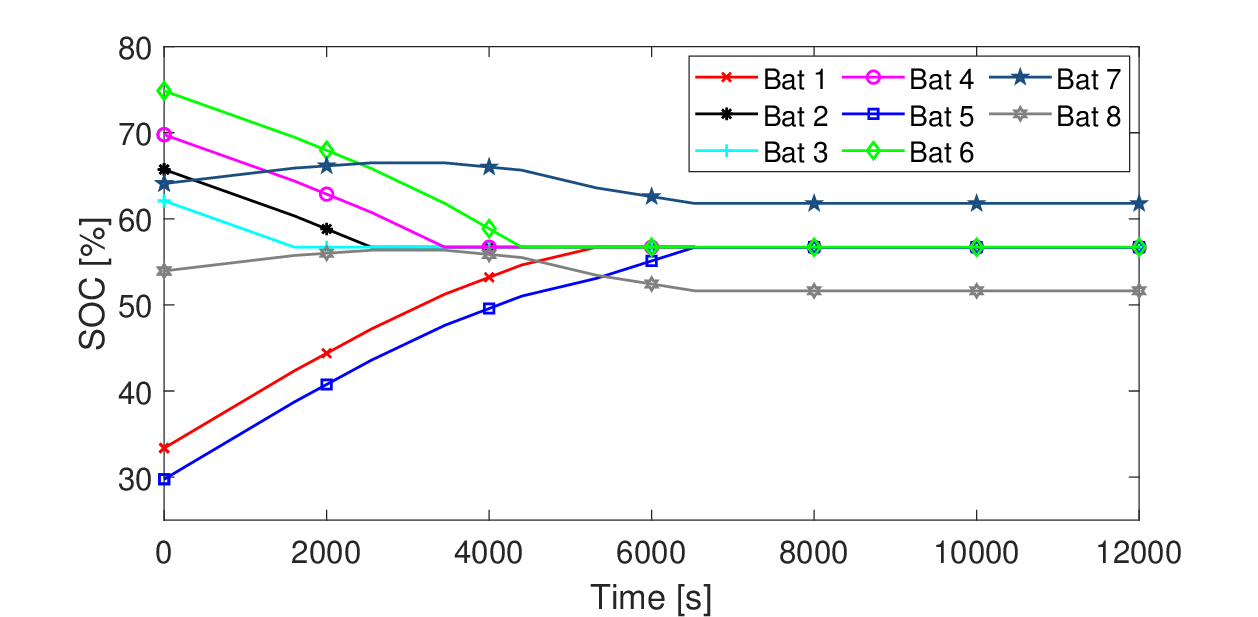}
		\end{minipage}
	}	
	\subfigure[]{
	\begin{minipage} {\linewidth}\label{CPCr4}
		\centering
		\includegraphics[width=0.9\linewidth, trim=1 1 20 10, clip]{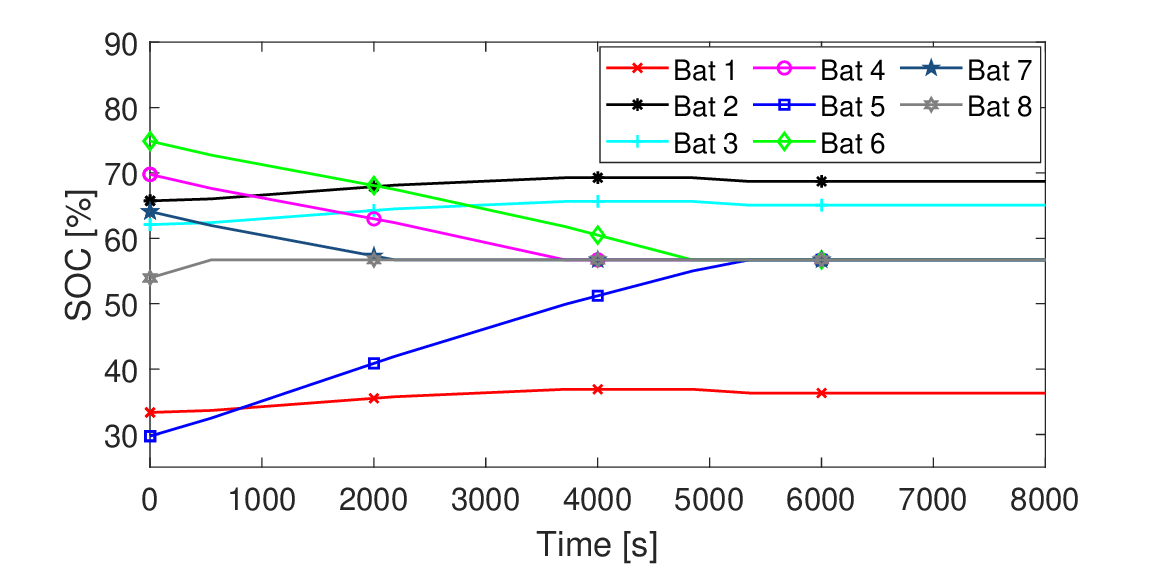}
	\end{minipage}
}	
	\caption{Simulation results for CPC equalization systems: (a) traditional, (b)  without $e_n$, (c) without $e_{n-1}$ and $e_n$, (d) without $e_{1}$, $e_2$, and $e_3$.}
	\label{Re1}
\end{figure}

\begin{figure}[!htbp]
	\centering		
	\subfigure[]{
		\begin{minipage} {\linewidth}\label{MCPCo}
			\centering
			\includegraphics[width=0.9\linewidth, trim=1 1 20 10, clip]{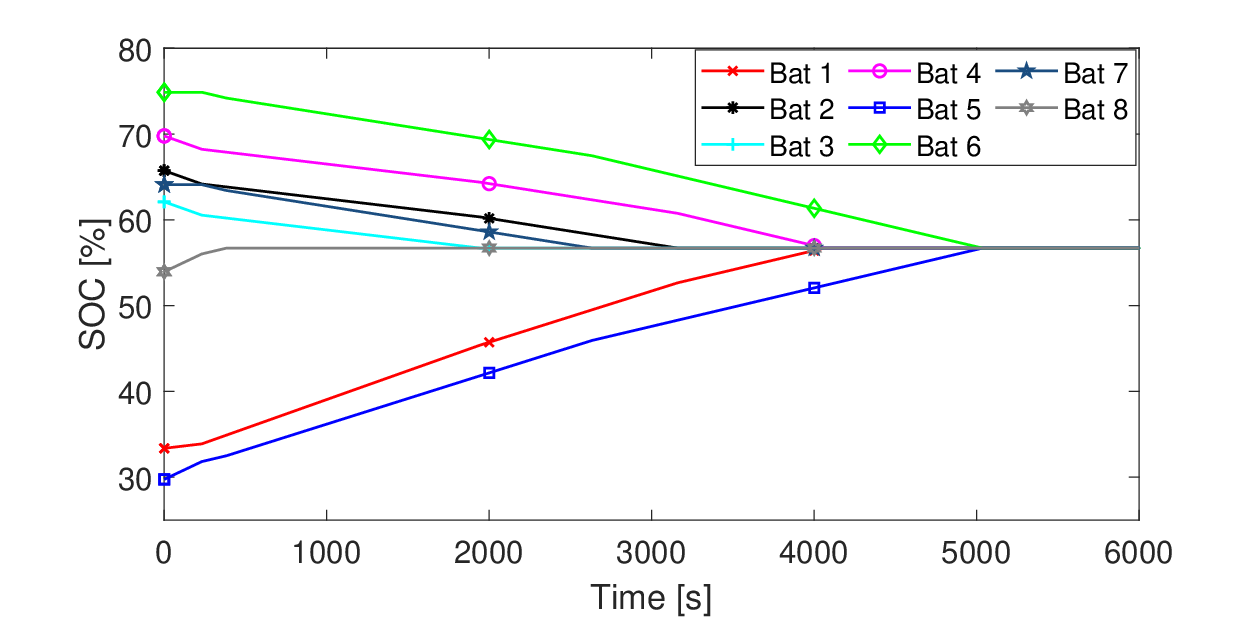}
		\end{minipage}
	}	
	\subfigure[]{
		\begin{minipage}{\linewidth}\label{MCPC1u}
			\centering
			\includegraphics[width=0.9\linewidth, trim=1 1 20 10, clip]{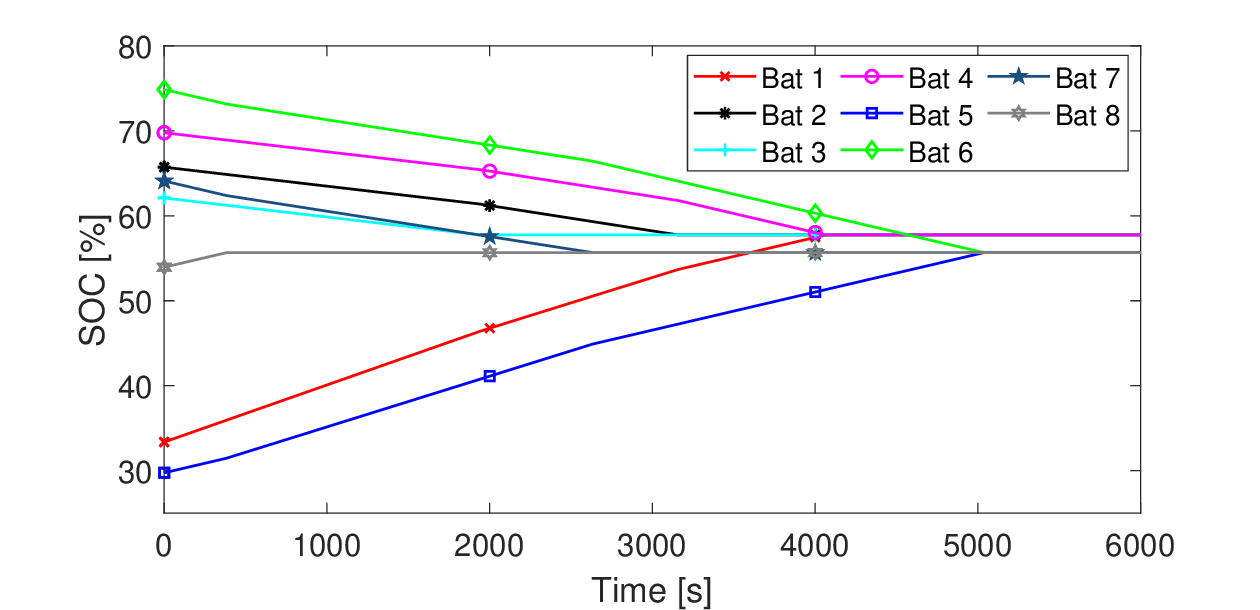}
		\end{minipage}
	}	
	\subfigure[]{
		\begin{minipage} {\linewidth}\label{MCPC2}
			\centering
			\includegraphics[width=0.9\linewidth, trim=1 1 20 10, clip]{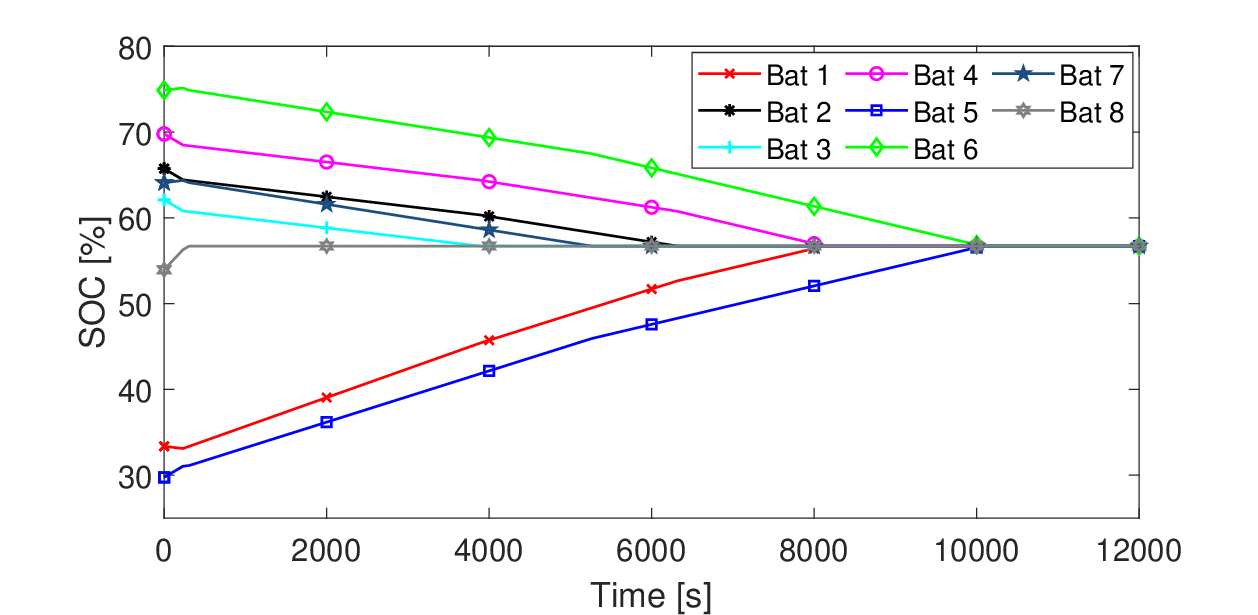}
		\end{minipage}
	}	
	\subfigure[]{
	\begin{minipage} {\linewidth}\label{MCPCr3}
		\centering
		\includegraphics[width=0.9\linewidth, trim=1 1 20 10, clip]{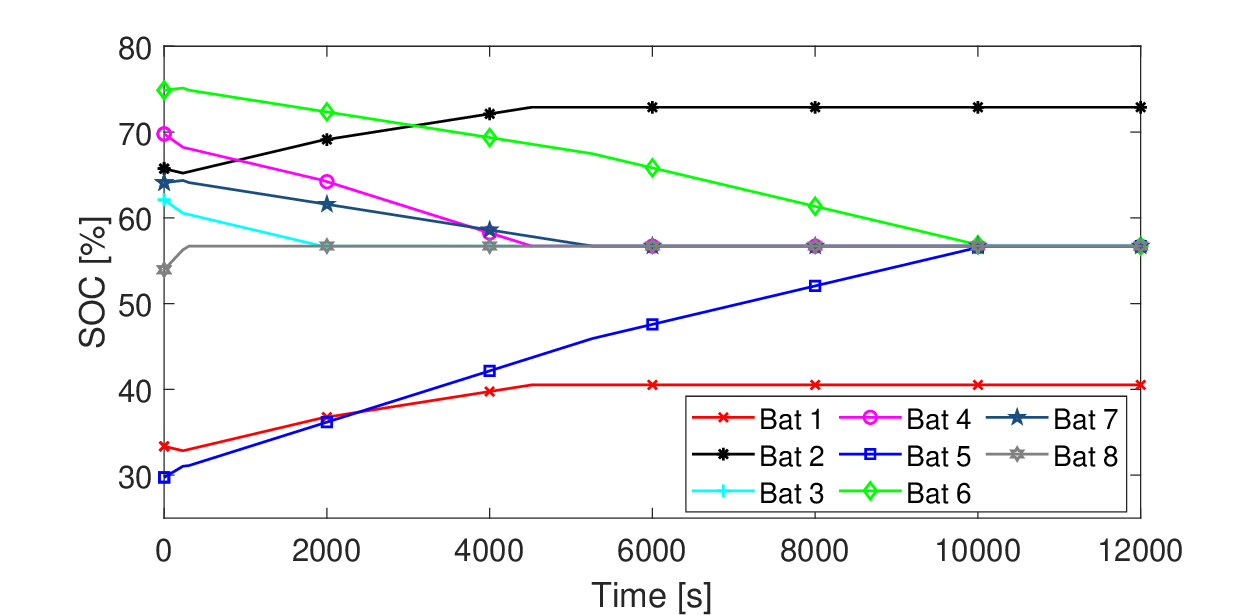}
	\end{minipage}
}	
	\caption{Simulation results for module-based CPC equalization systems: (a) traditional, (b)  without $e_1$, (c) without $e_2$ and $e_6$, (d) without $e_2$, $e_3$, and $e_6$.}
	\label{Re2}
\end{figure}

	\begin{table}[!htbp]
		\caption{Comparison results of equalization systems with different numbers of equalizers}
		\begin{center}
			\setlength{\tabcolsep}{1mm}{
				\begin{tabular}{cl|ccc}
					\hline 	
				 &   &  \multirow{2}{*}{\text{rank}(C)} & Number of & Equalization  \\
				  &   &  &  Equalizers & achieved? \\
					\hline
				\multirow{2}{*}{CPC }  & Traditional  & $n-1$ & $n$ & Yes \\
				  & Without $e_n$  & $n-1$ & $n-1$ & Yes \\
				($n=8$)  & Without $e_{n-1}$ and $e_n$  & $n-2$  & $n-2$ & No \\
			   	  & Without $e_1$, $e_2$, and $e_3$  & $n-3$  & $n-3$ &  No \\
				\hline 	
				\multirow{3}{*}{Module-based}  & Traditional  & $n-1$ & $n+1$ & Yes \\
				& Without $e_1$  & $n-2$ & $n$ & No  \\
			 \multirow{1}{*}{CPC}	& Without $e_2$ and $e_6$  & $n-1$ & $n-1$ & Yes \\
				($n=8$, $m=2$)	& Without $e_2$, $e_3$, and $e_6$  & $n-2$ & $n-2$ & No \\
					\hline
			\end{tabular}}
		\end{center}
		\label{LC}
	\end{table}

	\begin{table*}[!htbp]
		\caption{Average equalization time comparison for Six battery
			equalization systems}
		\begin{center}
			\setlength{\tabcolsep}{1mm}{
				\begin{tabular}{cc|cccccc}
					\hline 	
					&	&  Series-based CC &  Module-based CC & Layer-based CC & CPC &  Module-based CPC & Switch-based CPC\\
					\hline
					
					\multirow{2}{*}{$n=8$, $m=2$} & $\lambda_{n-1}(CC^T)$ & $0.1522$  & $0.5858$  &  $2$ & $1$ & $1$  & $0$\\
					& Average equalization time &  $4680.1~\rm{s}$  & $3562~\rm{s}$   & $2675.9~\rm{s}$ & $3350~\rm{s}$  & $3076~\rm{s}$   & $26501~\rm{s}$ \\
					\hline 	
					
					\multirow{2}{*}{$n=16$, $m=2$} & $\lambda_{n-1}(CC^T)$ & $0.0384$   & $0.1522$ & $2$  & $1$ & $1$  & $0$\\
					& Average equalization time & $6967.6~\rm{s}$ & $5443.3~\rm{s}$  & $2960.6~\rm{s}$ & $3699.1~\rm{s}$ & $3559.8~\rm{s}$ & $57912~\rm{s}$   \\
					\hline
					\multirow{2}{*}{$n=32$, $m=4$} & $\lambda_{n-1}(CC^T)$ & $0.0096$  &  $0.1522$ & $2$ & $1$ & $1$  & $0$ \\
					& Average equalization time &  $9585.3~\rm{s}$   & $5945.7~\rm{s}$  & $3079~\rm{s}$ & $3762.4~\rm{s}$  & $3585.1~\rm{s}$  &  $116340~\rm{s}$   \\
					\hline
					\multirow{2}{*}{$n=64$, $m=4$} & $\lambda_{n-1}(CC^T)$ & $0.0024$  & $0.0384$  & $2$  & $1$ & $1$ & $0$ \\
					& Average equalization time &  $11983~\rm{s}$  & $8003.3~\rm{s}$ & $3004.4~\rm{s}$ & $3667.5~\rm{s}$ & $3583~\rm{s}$ & $224080~\rm{s}$ \\
					\hline
							\multirow{2}{*}{$n=128$, $m=8$} & $\lambda_{n-1}(CC^T)$ & $0.0006$  &  $0.0384$ & $2$ & $1$ & $1$ & $0$  \\
					& Average equalization time &  $12740~\rm{s}$ & $7513.9~\rm{s}$ & $2800.9~\rm{s}$ & $3495.9~\rm{s}$ & $3403.3~\rm{s}$  & $421180~\rm{s}$ \\
					\hline
			\end{tabular}}
		\end{center}
		\label{LC2}
	\end{table*}

\subsection{Validation of the Minimum Required Number of Equalizers} 
To demonstrate Lemma~\ref{L1} regarding the minimum number of equalizers required to achieve equalization, simulations are conducted on a battery pack consisting of $8$ cells. The capacity of battery cells is set as $3.1\rm{Ah}$ and their initial SOCs are selected as $x(0)=[33.37\%, 65.73\%,  62.1\%, 69.78\%, 29.75\%, 74.87\%,$ $ 64.1\%, 53.95\%]^T$. For the CPC equalization system, there are $n=8$ equalizers as illustrated in Fig. \ref{combined6}(d), and the SOC responses are shown in Fig. \ref{CPCtr}. If we delete the equalizer $e_n$, the number of equalizers becomes $n-1$, and $\text{rank}(C)=n-1=7$. The corresponding battery equalization process is illustrated in Fig. \ref{CPCr2}, confirming that the equalization is still feasible. In contrast, upon removing the equalizers $e_{n-1}$ and $e_{n}$, or removing  $e_{1}$, $e_{2}$, and $e_3$, the number of equalizers becomes $n-2$ (namely, $\text{rank}(C)=6$) or $n-3$ ($\text{rank}(C)=5$), respectively. Consequently, these two equalization systems do not satisfy the necessary condition of controllability in Lemma~\ref{L1}. Figs.~\ref{CPCr3} and \ref{CPCr4} show the results of the corresponding SOC responses, demonstrating that the equalization cannot be completed.

Furthermore, simulation results are also given in Fig.~\ref{Re2} for the traditional module-based CPC equalization system, the CPC equalization systems without $e_1$, without $e_2$ and $e_6$, and without $e_2$, $e_3$, and $e_6$, respectively. The results summarized in TABLE~\ref{LC} demonstrate the efficiency of the necessary condition highlighted in Lemma~\ref{L1} and the minimum required number of equalizers for battery equalization systems.

\subsection{Equalization Time Comparison Results}
Based on the Monte Carlo method, more simulations are conducted under the umbrella of Algorithm~\ref{algo_AA} to compare the equalization time for battery equalization systems. The number of series-connected cells in the battery pack is selected from $\{ 8, 16, 32, 64, 128\}$, and the cells' initial SOCs are generated independently and randomly for $50000$ times from the uniform distribution $U(40\%,80\%)$. The equalization currents
are all set to $0.5~\rm{A}$. The tolerance bound is set to $\epsilon=0.1\%$.

\begin{figure}[!htbp]
	\centering		
	\subfigure[]{
		\begin{minipage} {\linewidth}\label{Sta8-5}
			\centering
			\includegraphics[width=0.9\linewidth, trim=1 1 20 10, clip]{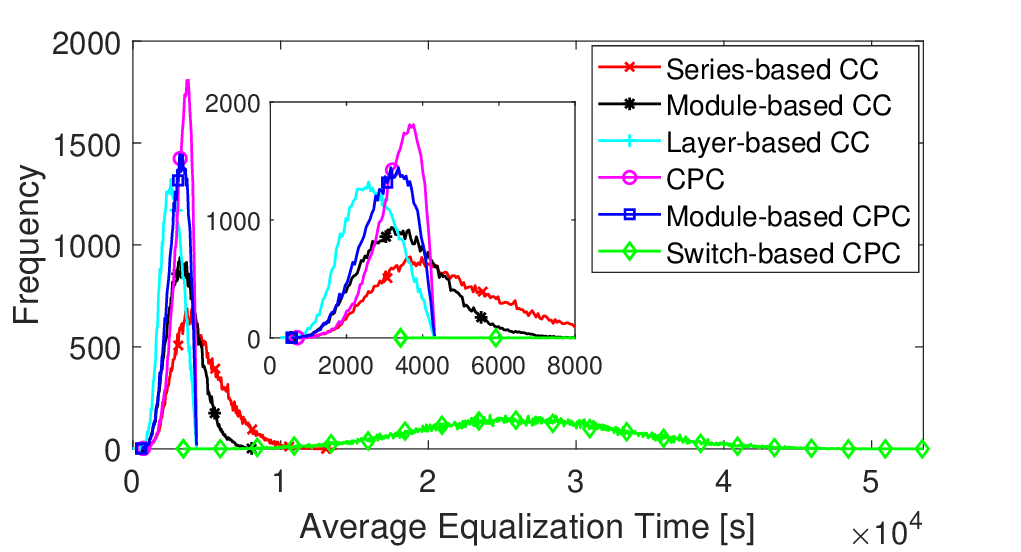}
		\end{minipage}
	}	
	\subfigure[]{
		\begin{minipage}{\linewidth}\label{Sta16-5}
			\centering
			\includegraphics[width=0.9\linewidth, trim=1 1 20 10, clip]{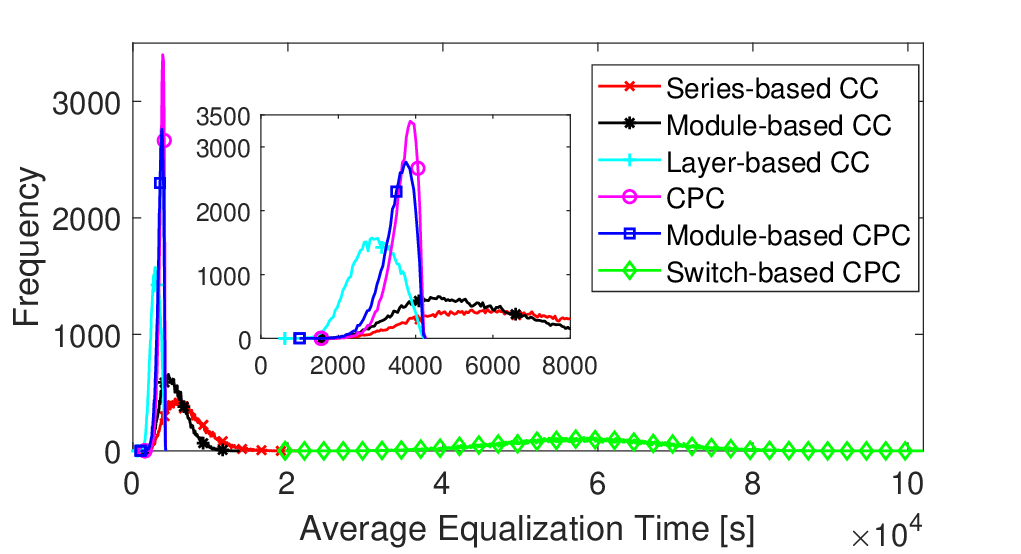}
		\end{minipage}
	}	
	\caption{Equalization time distribution of six battery equalization systems for the battery pack with (a) $8$ and (b) $16$ cells.}
	\label{Re3}
\end{figure}

As indicated in (\ref{43}), the equalization time is inversely correlated with the second smallest eigenvalue of $CC^T$ in an equalization system, so we calculate $\lambda_{n-1}(CC^T)$ for each system. TABLE~\ref{LC2} includes the values of $\lambda_{n-1}(CC^T)$ and the average equalization times resulting from Monte Carlo simulations, for all the six equalization systems. By sorting all these systems in descending order based on $\lambda_{n-1}(CC^T)$, the obtained sequence is: layer-based CC, module-based CPC, CPC, module-based CC, series-based CC, and switch-based CPC equalization systems, respectively. This order of $\lambda_{n-1}(CC^T)$ for these equalization systems is consistent with the ranking of their average equalization times derived from simulations. It shows that the layer-based CC system has the fastest equalization on average, and the switch-based CPC system is the slowest. Hence, the power of Lemma~\ref{L2} is numerically demonstrated. The above conclusions can also be reflected from the frequency distributions of equalization time for the battery pack with $8$ and $16$ cells, as shown in Fig.~\ref{Re3}.

In addition, the relationship between $\lambda_{n-1}(CC^T)$ and equalization time can also be used for determining the number of modules $m$ in the battery pack equipped with a module-based CC equalization structure. Based on the results in TABLE~\ref{LC22}, it can be concluded that increasing $m$ within a certain range will enlarge $\lambda_{n-1}(CC^T)$ and shorten the equalization time.

	\begin{table}[!htbp]
		\caption{Average equalization time comparison for battery equalization systems with different number of modules}
		\begin{center}
			\setlength{\tabcolsep}{1mm}{
				\begin{tabular}{c|cc}
					\hline 	
					 &  $\lambda_{n-1}(CC^T)$ & Average equalization time   \\
				
					\hline
				  $n=64$, $m=2$  & $0.0096$ & $10026~\rm{s}$  \\
					 $n=64$, $m=4$  & $0.0384$ & $8003.3~\rm{s}$   \\
				   $n=64$, $m=8$  & $0.1522$  & $6045.6~\rm{s}$   \\
					\hline 	
					  $n=128$, $m=4$  & $0.0096$ &  $9413.7~\rm{s}$ \\
					 $n=128$, $m=8$  & $0.0384$ &   $7513.9~\rm{s}$ \\
					   $n=128$, $m=16$  & $0.1522$ &   $5663.7~\rm{s}$   \\
					\hline
			\end{tabular}}
		\end{center}
		\label{LC22}
	\end{table}

\section{Conclusions}
This study introduced an innovative hypergraph-based modeling approach for active battery equalization systems, which reveals the inherent relationship between cells and equalizers. Based on the developed unified model and its controllability analysis, the minimum required number of equalizers has been derived for effective battery balancing. The identified correlation between equalization time and the second smallest eigenvalue of the equalization system's Laplacian matrix can significantly simplify the comparison and optimized design of various equalization systems, eliminating the need for extensive experiments or simulations to derive equalization times. Furthermore, this correlation provides insights for optimizing the equalization time. The proposed modeling framework paves the way for future research on the refinement of control strategies and the advancement of battery equalization technologies. In the future, we plan to extend the proposed unified model, incorporating factors such as the state-of-health information of battery cells to enhance its applicability and functionality for advanced battery management.

\ifCLASSOPTIONcaptionsoff
\newpage
\fi

\bibliographystyle{IEEEtran}
\bibliography{mybibfile}

\end{document}